\documentclass[mathptm]{jcm}
\usepackage{comment,enumerate,changebar}
\usepackage{epsf}
\usepackage{amssymb}
\usepackage{amsmath,amsthm}
\usepackage{amssymb}
\usepackage{latexsym}
\usepackage{psfrag}
\usepackage{graphicx,subfigure}
\usepackage[usenames,dvipsnames]{color}
\newtheorem{theo}{Theorem}[section]
\newtheorem{prop}[theo]{Proposition}
\newtheorem{lemma}[theo]{Lemma}
\newtheorem{defn}[theo]{Definition}
\newtheorem{exam}[theo]{Example}
\newtheorem{examples}[theo]{Examples}
\newtheorem{rem}[theo]{Remark}

\newtheorem{exercise}[theo]{Exercise}
\newtheorem{exercises}[theo]{Exercises}

\newtheorem{corollary}[theo]{Corollary}

\newenvironment{remark}{\begin{rem} \rm}{\end{rem}}
\newenvironment{ex}{\begin{exam}\rm}{\end{exam}}

\newcommand{\nc}{\newcommand}

\nc{\RR}{{\mathbb R}} %% Real Numbers
\nc{\ZZ}{{\mathbb Z}} %% Integers
\nc{\NN}{{\mathbb N}} %% Natural Numbers
\nc{\nat}{\NN}
\nc{\QQ}{{\mathbb Q}} %% Rational Numbers
\nc{\HH}{{\mathbb H}} %% Hyperbolic Space
\nc{\PP}{{\mathbb P}} %% Projective Space
\nc{\CC}{{\mathbb C}} %% Complex numbers
\nc{\co}{\CC}
\nc{\idgp}{\e} %% Identity Of A Group

\nc{\Ima}{\mbox{im}} %% Image of a map
\nc{\im}{\mbox{{\rm Im}}} %% Image of a map
\nc{\Supp}{\mbox{Supp}} %% Support of a map
\nc{\kr}{\mbox{ker}} %% Kernel of a map
\nc{\coker}{\mbox{coker}} %% Cokernel of a map

\nc{\Sym}{\mbox{Sym}} %% Symmetric Group
\nc{\GL}{\mbox{GL}} %% General Linear Group
\nc{\SL}{\mbox{SL}} %% Special Linear Group
\nc{\PSL}{\mbox{PSL}} %% Projective Special Linear Group
\nc{\End}{\mbox{End}} %% Group of endomorphisms
\nc{\Aut}{\mbox{Aut}} %% Group of automorphisms
\nc{\Out}{\mbox{Out}} %% Group of outer automorphisms
\nc{\Inn}{\mbox{Inn}} %% Group of inner automorphisms
\nc{\Hom}{\mbox{Hom}} %% Hom functor
\nc{\Homeo}{\mbox{Homeo}} %% Group of homeomorphisms
\nc{\Stab}{\mbox{Stab}} %% Stabilizer
\nc{\Orb}{\mbox{Orb}} %% Orbit
\nc{\id}{\mbox{id}} %% Identity
\nc{\Conj}{\mbox{Conj}} %% Conjugacy Class
\nc{\FP}{\mbox{FP}} %% Higher Finiteness Condition

\nc{\val}{\mbox{val}} %% Valency
\nc{\str}{\mbox{star}} %% Lower case star
\nc{\Star}{\mbox{Star}} %% Upper case star
\nc{\geom}{\mbox{geom}} %% Geometric realisation

\nc{\tr}{\mbox{tr}} %% Lower case trace
\nc{\Tr}{\mbox{Tr}} %% Upper case trace

\nc{\defi}{\mbox{def}} %% For use in =def
\nc{\e}{\epsilon}
\nc{\ep}{\varepsilon} %% Shorthand for epsilon
\nc{\half}{\frac{1}{2}}
\nc{\bl}{\vspace{2mm}\\} %% Blank Line
\nc{\lte}{\leqslant} %% Less than or equal to
\nc{\gte}{\geqslant} %% Greater than or equal to
\nc{\ltec}{\preceq} %% lte curly (for posets)
\nc{\gtec}{\succeq} %% gte curly (for posets)
\nc{\ltc}{\prec} %% less than curly
\nc{\gtc}{\succ} %% greater than curly
\nc{\pmd}{^{\prime}} %% Primed
\nc{\pmmd}{^{\prime\prime}} %% Double primed
\nc{\ppmd}{^{\prime\prime}} %% Alternative abbreviation
\nc{\ra}{\rightarrow} %% Arrow pointing right, e.g. definition of a map
\nc{\bks}{\backslash}
\nc{\rank}{\mbox{rank}}
\nc{\Min}{\mbox{Min}} %% Capital min
\nc{\lcm}{\mbox{lcm}} %% Lowest common multiple
\nc{\modn}{\mbox{ }\mod \mbox{ } n}
\nc{\Hull}{\mbox{Hull}}
\nc{\sgn}{\mbox{sgn}}
\nc{\hcf}{\mbox{hcf}} % highest common factor
\nc{\ann}{\mbox{Ann}} %Annihilator
\nc{\stab}{\mbox{Stab}} %Stabilizer

\nc{\diff}{\mbox{diff}}
\nc{\slice}{\mbox{slice}}
\nc{\diam}{\mbox{diam}}
\nc{\ce}{\mbox{ce}}
\nc{\st}{\mbox{st}}
\nc{\Ca}{\mbox{Ca}}
\nc{\con}{\mbox{con}}
\nc{\State}{\mbox{State}}
\nc{\DState}{\mbox{DState}}
\nc{\exce}{\mbox{exc}}

\nc{\cone}{\mbox{Con}}
\nc{\Con}{\cone}
\nc{\asdim}{\mbox{asdim}}
\nc{\dis}{\mbox{dis}}
\nc{\Cc}{\mbox{Cc}}
\nc{\PD}{\mbox{PD}}
\nc{\cd}{\mbox{cd}}
\nc{\chr}{\mbox{char}}
\nc{\Diffeos}{\mbox{Diffeos}}
\nc{\Fix}{\mbox{Fix}}
\nc{\Fx}{\mbox{Fix}}
\nc{\Eq}{\mbox{Eq}}
\nc{\Ends}{\mbox{ends}}
\nc{\Vol}{\mbox{Vol}}
\nc{\Mon}{\mbox{Mon}}
\nc{\WCH}{\mbox{WCH}}

\nc{\Iso}{\mbox{Iso}} %% Isometry group of a metric space
\nc{\Int}{\mbox{Int}}
\nc{\inter}{\mbox{Int}}

\nc{\OS}{\mbox{OS}}
\nc{\IS}{\mbox{IS}}
\nc{\two}{{\bf 2}}
\nc{\tens}{\otimes}
\nc{\zvect}{\underline{0}}
\nc{\QFT}{\mbox{QFT}}
\def \ket#1{|#1\rangle}

\def \set#1{\{#1\}}

\def \ip#1{\langle#1\rangle}

\def \eqdefn{=^{\rm defn}}

\begin{document}
%\shownotes
\begin{comment}
\begin{center}
{\large \bf Extending the Promise of the Deutsch--Jozsa--H\o yer Algorithm for Finite Groups}
\end{center}
\begin{center}
{\large Michael Batty, Samuel L. Braunstein and Andrew J. Duncan}
\end{center}
\begin{center}
{\large \today}
\end{center}

\begin{abstract}
H\o yer has given a generalisation of the Deutsch--Jozsa algorithm which uses the Fourier transform
on a group $G$ which is (in general) non-Abelian. His algorithm distinguishes between functions 
which are either perfectly balanced ($m$-to-one) or constant, with certainty, and using a single 
quantum query. Here, we show that this algorithm (which we call the Deutsch--Jozsa--H\o yer 
algorithm) can in fact deal with a broader range of promises, which we define in terms of the 
irreducible representations of $G$.
\end{abstract}
\end{comment}
%%%%%%%%%%%%%%%%%%%%%%%%%%% jcm stuff
\title
  [Extending the Deutsch--Jozsa--H\o yer Algorithm]         % |Optional short form, for the running head|
  { Extending the Promise of the Deutsch--Jozsa--H\o yer Algorithm for Finite Groups} % |Main Title|
\author
{ 
  Michael Batty
}
\email
{ 
  Michael.Batty@ncl.ac.uk
}
\author
{
  Andrew J. Duncan
}
\email
{
  A.Duncan@ncl.ac.uk
}
\address
{
  Department of Mathematics,\\    
  School Of Mathematics and Statistics,\\
  Merz Court,\\ University of Newcastle upon Tyne,\\
  Newcastle upon Tyne,\\NE1 7RU,\\ United Kingdom.
}
% |A 2nd author|
\author{Samuel L. Braunstein}
\email{schmuel@cs.york.ac.uk}
\address{
 Department of Computer Science, \\
        University of York, \\
        York, \\
        YO10 5DD, \\
        United Kingdom.         
}
% |A third author|
% |Math Reviews Classification numbers|

% |Abstract comes before maketitle, as in the AMS classes|
\begin{abstract}
% |Abstract text|
H\o yer has given a generalisation of the Deutsch--Jozsa algorithm which uses the Fourier transform
on a group $G$ which is (in general) non-Abelian. His algorithm distinguishes between functions 
which are either perfectly balanced ($m$-to-one) or constant, with certainty, and using a single 
quantum query. Here, we show that this algorithm (which we call the Deutsch--Jozsa--H\o yer 
algorithm) can in fact deal with a broader range of promises, which we define in terms of the 
irreducible representations of $G$.
\end{abstract}

\maketitle

%%%%%%%%%%%%%%%%%%%%%%%%%%%
\section{Introduction}
Recall that a function $f:\{0,1\}^n\ra\{0,1\}$ is called {\em balanced} if
$|f^{-1}(0)|=|f^{-1}(1)|=2^{n-1}$.
Deutsch's algorithm \cite{deutsch} distinguishes between constant and balanced functions
from $\{0,1\}$ to itself using a single quantum query, whereas classically two queries are
required. A function from $\{0,1\}$ to itself is always either balanced or constant. However,
to generalise to functions $\{0,1\}^n\ra\{0,1\}$, Deutsch and Jozsa \cite{deutschjozsa} realised
that we must restrict the class of functions condsidered. They showed that we 
can distinguish between constant and balanced functions $\{0,1\}^n\ra\{0,1\}$, again in a single
query, but if we are given a function $\{0,1\}^n\ra\{0,1\}$ which is neither constant nor 
balanced, then we can't deduce anything from the output of the quantum circuit. Thus, we must
be {\em promised} that the function is either constant or balanced; then we can use the circuit
to deduce something. 

It was first realised by H\o yer that the mathematics  underlying the Deutsch--Jozsa 
algorithm is group-theoretic in nature. In \cite{hoyer2}, he remarks that if we replace the 
discrete Fourier transform on $\ZZ_2^n$ employed in the Deutsch--Jozsa algorithm by the Fourier
transform on an arbitrary finite group, then we can distinguish between constant and perfectly 
balanced functions. In this paper we show that the range of functions which can be distinguished
is broader than this, provided that we make corresponding promises. These promises are 
representation-theoretic in nature, further reflecting the role played by finite groups in the 
Deutsch--Jozsa circuit.

The definitions of the types of functions we consider seem at first
sight somewhat technical and perhaps unnatural. However, given a map
$f:X\ra H$, where $H$ is a finite group we
associate an element $r$ of the integral group ring $\ZZ H$ to $f$ in
such a way the promise on the function becomes a promise on the
element $r$; namely that $r$ lies in one of two subsets of $\ZZ H$
which have natural and straightforward descriptions (see Section
\ref{section:ab}).

In the case of functions $f:X\ra A$ where $A$ is an Abelian group our 
promises can be described in terms of a polynomial $P_f$ associated to
$f$. In fact, as we show in Section \ref{section:ab}, our
representation-theoretic promise is
equivalent to the promise that $P_f$ is either monomial or is divisble
by the $n$th cyclotomic polynomial, where $n=|A|$ (see Section
\ref{section:appendix}). If $n$ has at most $2$ distinct prime
divisors then this gives rise to a further characterisation of the
promise on $f$ in terms of certain subgroups of $A$. 

The paper is structured as follows.  In Sections \ref{section:oracles} and \ref{section:qft} we
describe quantum oracles for group multiplication and give a
definition of the quantum Fourier transform, convenient for
our purposes. In Section \ref{section:ab} we define our representation-theoretic
versions of constant and balanced functions, characterise these types
of function in terms of the integral group ring, discuss the case
where the codomain is Abelian and give examples where the codomain
is non-Abelian. Section \ref{section:algorithm} explains how the Deutsch--Jozsa circuit
is used to distinguish between constant and balanced
functions, of this kind. Section \ref{section:cases} lists various other algorithms
which are special cases of our algorithm. 
Appendix 1, Section \ref{section:appendix0}, gives a brief introduction to
group representation theory and the ``Weyl trick''.
Appendix 2, Section
\ref{section:appendix},
covers the number theory used in Section \ref{section:ab}.
\begin{comment}
In an earlier draft of this paper, we gave a logarithmic-time algorithm to distinguish between
types of functions $f:\{0,1\}^n\ra\{0,1\}^n$. We are grateful to an anonymous referee for 
pointing out that this can be achieved by a logarithmic number of applications of the 
{\em standard} Deutsch--Jozsa algorithm. This is because the group $\ZZ_2^n$ can be expressed as
a direct product of many smaller groups. In an arbitrary finite group no such direct decomposition
will exist. We therefore believe that the promises we give in this paper are genuinely more
general than the Deutsch--Jozsa promise.
\end{comment}

\section{A Quantum Oracle for Group Multiplication}
\label{section:oracles}
The following definition generalises the notion of a qubit.
\begin{defn} Let $X$ be a finite set. A {\em quX} is a complex vector space spanned by
$\set{\ket{x}\mid x\in X}$. 
\end{defn}
\noindent
For example, a qubit is a qu$\set{0,1}$.

Suppose that $G$ is a finite group and $X$ is a finite set. Write $\Sym(X)$ for the group of 
permutations of $X$. 
Suppose that we are also given a function (not necessarily a homomorphism)
$\theta:G\ra\Sym(X)$.
Define a map $\phi:G\times X\ra G\times X$
by the rule $\phi:(g,x)\mapsto (g,[\theta(g)](x))$. 
If $\phi(g,x)=\phi(h,y)$ then $g=h$ and
$[\theta(g)](x)=[\theta(g)](y)$, in which case $x=y$, as $\theta(g)$
is a permutation; that is $\phi$ is an injection, and as $G\times X$
is finite, it is a bijection.
Now suppose that we have
a quantum system $\CC^{G\times X}\cong\CC^G\tens\CC^X$ 
comprising two quantum registers, a qu$G$ and a qu$X$. 
Then there is a unitary map
$U$ which permutes the basis states of this system:
\[ U:\ket{g,x}\mapsto\ket{g,[\theta(g)](x)}. \]
In particular consider the following 
case.
 Suppose that $X$ is a group $H$ and $f:G\ra H$ is a function (not necessarily a homomorphism).
Define $[\theta(g)]h=f(g)h$. Then \[ U:\ket{g,x}\mapsto\ket{g,f(g)h}, \]
and we say that $U$ is the $H$-{\em multiplication oracle} for the function $f:G\ra H$.
For example, suppose that $G\cong (Z_2)^n$ and $H\cong (Z_2)^m$. Then we recover the usual
exclusive-OR oracle 
\[ U:\ket{x,y}\mapsto\ket{x,y\oplus f(x)}. \]
\begin{comment}
two cases.
\begin{enumerate}
\item \label{item:item1} Suppose that $G$ {\em acts} on $X$ via $[\theta(g)](x)=gx$ (i.e. $\theta$ {\em is} a homomorphism). 
If $\phi(g,x)=\phi(h,y)$ then $gx=gy$. Thus $g^{-1}(gx)=g^{-1}(gy)$ which gives $(g^{-1}g)x=(g^{-1}g)y$, i.e. $\idgp_Gx=\idgp_Gy$ which
gives $x=y$. Then  \[ U:\ket{g,x}\mapsto\ket{g,gx}, \]
and we say that $U$ is a {\em quantum oracle} for the action of $G$ on $X$.
\item \label{item:item2} Suppose that $X$ is a group $H$ and $f:G\ra H$ is a function (not necessarily a homomorphism).
Define $[\theta(g)]h=f(g)h$. Then \[ U:\ket{g,x}\mapsto\ket{g,f(g)h}, \]
and we say that $U$ is the $H$-{\em multiplication oracle} for the function $f:G\ra H$.
For example, suppose that $G\cong (Z_2)^n$ and $H\cong (Z_2)^m$. Then we recover the usual
exclusive-OR oracle 
\[ U:\ket{x,y}\mapsto\ket{x,y\oplus f(x)}. \]
\end{enumerate}
\end{comment}
%%%%%%%%%%%
%%%%%%%%%%%%%%%
\section{Representations and Non-Abelian Fourier Transforms}\label{section:qft}
\subsection{Irreducible Representations and the Quantum Fourier Transform}
\label{subsection:reps}
The quantum Fourier transform is an essential subroutine in nearly all the
quantum algortithms developed to date. First we define the transform and then 
briefly discuss its implementation.
%We now describe the quantum Fourier transform ${\cal F}_G$ on $G$. 
%For the reader unfamiliar with group representation theory, a brief introduction is given
%in section~\ref{section:appendix0}.
Recall that every finite group $G$ has only finitely many irreducible representations (see Section~\ref{section:appendix0}). By
application of the ``Weyl trick'' (Section~\ref{section:appendix0}), we may convert any finite dimensional representation $\rho$ of $G$ into an 
equivalent unitary representation $\rho\pmd$, and if $\rho$ is irreducible then so is its 
unitarization $\rho\pmd$ (since it is equivalent to $\rho$).
Let the (unitarized) irreducible representations of the finite group $G$ be $\rho^1,\ldots,\rho^r$
(from now on we will omit the primes).
Then these representations are used
to define the quantum Fourier transform on $G$. It is well known that $\sum_{j=1}^r(\dim \rho^j)^2=|G|$ (see e.g. \cite{fultonharris}). Let
\[ R=\set{(i,j,k)\mid 1\lte i,j \lte\dim \rho^k, 1\lte k\lte r}. \]
Then $|R|=|G|$ and we may specify a (non-canonical) bijection $\beta:G\ra R$. 
Writing $\idgp_G$ for the identity element of $G$, suppose also that
$\beta(\idgp_G)=(1,1,1)$ and that $\rho^1$ is the trivial representation. (This will aid calculations later.) 
Note that every matrix entry $\rho^k_{i,j}$ of an irreducible representation $\rho^k$ is a function from $G$ to $\CC$,
since the matrices in the representation vary over $G$. 
For ease of notation, write $\beta(g)=(i_g,j_g,k_g)$ and then write $\rho^g$ for the function from $G$ to $\CC$ defined by
\[ \rho^g(g\pmd)=\rho_{i_g,j_g}^{k_g}(g\pmd)\mbox{ for all $g\pmd\in G$}, \]
the $(i_g,j_g)^{\rm th}$ matrix entry of the $k_g^{\rm th}$ irreducible representation. Also, write $\dim(g)$ for
$\dim(\rho^{k_g})$. Then the Schur orthogonality relations (see \cite{terras}, p.251, Theorem 1. (1) and (2)) tell us
that 
\begin{equation*}%\label{eqn:sor} 
\ip{\rho^{g_1},\rho^{g_2}}\eqdefn \sum_{g\pmd\in G}\rho^{g_1}(g\pmd)\overline{\rho^{g_2}(g\pmd)}
=\left\{\begin{array}{l}\frac{|G|}{\dim(g)}\mbox{ if $g_1=g_2=g$}\\
0\mbox{ otherwise}\end{array}\right., 
\end{equation*}
which is to say that $\{\rho^g\}_{g\in G}$ is an orthogonal basis for $L^2(G)$, the inner product space of the functions
$f:G\ra\CC$ under pointwise addition, scalar multiplication and the above inner product. If we define
$\tau^g_G=\rho^g.\sqrt{\frac{\dim g}{|G|}}$ then 
\begin{equation}\label{eqn:sor} 
\ip{\tau^{g_1},\tau^{g_2}}
=\left\{\begin{array}{l}
1 \mbox{ if $g_1=g_2$}\\
0\mbox{ otherwise}\end{array}\right., 
\end{equation}
so
$\{\tau_G^g\}_{g\in G}$ is an ortho{\em normal} basis.
Note that in particular we have
\begin{equation}\label{eqn:tauidentity}
\tau_G^{\idgp_G}=\frac{1}{\sqrt{|G|}}. 
\end{equation}
%$\begin{equation}\label{eqn:tauproduct}
%\tau_G^g(ab) = \left[\sqrt{\frac{|G|}{\dim g}}\right]\tau_G^g(a)\tau_G^g(b).
%\end{equation}

%\subsection{The Quantum Fourier Transform}
We define the {\em quantum Fourier transform on} $G$ (with respect to the bijection $\beta$ which is suppressed in the notation) 
to be the unitary map 
${\cal F}_G^{\phantom{\dagger}}:\CC^G\ra\CC^G$ defined for all $g\in G$ by
\[ {\cal F}_G^{\phantom{\dagger}}\ket{g}=\sum_{g\pmd\in G}\tau_G^{g}(g\pmd)\ket{g\pmd}.
\]
The conjugate transpose of ${\cal F}_G$ is given by
\[{\cal F}_{G}^{\dagger}\ket{g}=\sum_{g\pmd\in G}\overline{\tau_G^{g\pmd}(g)}\ket{g\pmd}.\]
That is, the matrix of ${\cal F}_G$ is given by $({\cal F}_G)_{g,g\pmd}=\tau_G^g(g\pmd)$ and
the matrix of ${\cal F}_G^{\dagger}$ is given by $({\cal F}_G^{\dagger})_{g,g\pmd}=\overline{\tau_G^{g\pmd}(g)}$.
We have
\begin{eqnarray*}
{\cal F}_G^{\dagger}{\cal F}_G^{\phantom{\dagger}}\ket{g} & = & 
{\cal F}_G^{\dagger}\sum_{g\pmd\in G}\tau_G^g(g\pmd)\ket{g\pmd} \\ 
& = &
\sum_{g\pmd,g\pmmd\in G}\tau_G^g(g\pmd)
\overline{\tau_G^{g\pmmd}(g\pmd)}\ket{g\pmmd} \\
& = & \sum_{g\pmmd\in G}\delta_{g,g\pmmd}\ket{g\pmmd} \mbox{ (by (\ref{eqn:sor}))} \\
& = & \ket{g}.
\end{eqnarray*}
This further implies that 
\begin{equation}\label{eqn:ffdagger}
{\cal F}_G^{\phantom{\dagger}}{\cal F}_G^{\dagger}=I,
\end{equation}
since
if $AB=I$ for any square (finite-dimensional) 
matrices $A$ and $B$ of the same size, then it follows
that we also have $BA=I$.   
Thus ${\cal F}_G^{\phantom{\dagger}}$ is unitary.
%\subsection{The Quantum Character Transform}
%A related construction is that of the {\em quantum character transform}.
%For an irreducible character $\chi$ of $G$ we define 
%\[ X\ket{\chi}=\frac{1}{\sqrt{|G|}}\sum_{g\in G}\chi(g)\ket{g}\in\CC^{G} \]
%We also define, given $g\in G$, 
%\[ X^{\dagger}\ket{g}=\frac{1}{\sqrt{|G|}}\sum_{\chi\in\hat{G}}\overline{\chi(g)}\ket{\chi}
%\in\CC^{\hat{G}}. \]
%Restricting $X^{\dagger}$ to $X(\CC^{\hat{G}})$, it is easily checked (using orthogonality 
%of characters) that for all $\chi\in\hat{G}$ we have
%$X^{\dagger}X\ket{\chi}=\ket{\chi}$ and it follows that $X$ is a unitary transformation.

The quantum Fourier transform  can be efficiently 
implemented in the case where $G$ is a finitely generated Abelian
 group using the classical
``Fast Fourier Transform'' \cite{shor}, \cite{coppersmith}. 
 Note that by an efficient
algorithm is meant one which runs in time polynomial in $\log(|G|)$.
 It is still unknown whether or not there is an efficient algorithm for the 
quantum Fourier transform over an arbitrary
finite group, although such algorithms exist in many cases
(\cite{beals}, \cite{hoyer}, \cite{maslenrockmore}, \cite{eh}, \cite{rb},
\cite{prb}).
In \cite{mrr} Moore, Rockmore and Russell survey and extend the results cited above, 
describing efficient algorithms 
for the quantum Fourier transform in several classes
of groups including  the symmetric groups $S_n$; wreath products $K\wr S_n$, where
$|K|$ is bounded  by a polynomial in $n$; metacyclic groups (a group $G$ 
is metacyclic if it has a cyclic normal subgroup $K$ such that $G/K$ is cyclic)
and metabelian (a group $G$ 
is metabelian if it has an Abelian normal subgroup $K$ such that $G/K$ is Abelian).
In particular all the groups in the examples of Section \ref{sub:nonabelian} below 
are covered by these classes.

\section{Generalisations of Constant and Balanced Functions}\label{section:ab}
Let $X$ be a finite set, let $H$ be a finite group and let $f:X\ra H$ be a function.
We assume the notation from the previous section for 
representations of finite groups. When we wish to apply the quantum Fourier
transform  to the set $X$ we regard it as the cyclic group $\ZZ_n$,
where $|X|=n$.
%\subsection{Generalised Constant and Balanced Functions}\label{section:genconst}
\begin{defn}\label{defn:const} Let $\rho^k$ be an irreducible (unitary) representation of $H$. Let 
$n=\dim \rho^k$ and suppose that $i\in \{1,\ldots,n\}$. We say that $f$ is
$\rho_i^k${\em -constant} if for each $r\in\{1,\ldots,n\}$ there exists a constant
$c_r\in\CC$ such that for all $g\in X$ we have $\tau_{i,r}^k(f(g))=c_r$.
\end{defn}
%%%%%%%%%%%%%%%%%%%%%%5
If $\chi$ is a linear ($1$-dimensional) representation 
%(i.e. a $1$-dimensional  character) 
of $H$ then we may simply refer to $f$ being ``$\chi$-constant''.
Recall that linear representations coincide with their characters and
that the set of
linear representations of $H$ forms a group. In the case of an Abelian
group the irreducible representations are all linear and we denote
this group $\hat H$ (see Section \ref{section:ab} below). 
If $H$ is an Abelian group and $h\in H$ then we adopt the practice of referring 
to ``$h$-balanced'', meaning $\chi$-balanced, where $\chi$ is the character 
corresponding to $h$ under the canonical isomorphism between $H$ and its group of characters $\hat{H}$.
Note that if $\chi_0$ is the trivial character of $H$ then every function from $X$ to $H$ is
$\chi_0$-constant, so we normally only consider $\chi$-constant functions for non-trivial 
characters $\chi$. 

Let $H=\ZZ_n$ and let the set of irreducible characters of $H$ be 
$\{\chi_k\}_{k\in H}$, where 
\[ \chi_k(x)=\exp\left({\frac{2\pi i k x}{n}}\right), \textrm{ for } x\in H.
\]
Then $f:X\ra H$ is $k$-constant if and only if there exists a  complex number 
$e^{i\theta}$ ($\theta\in\RR$) such that for all $s\in X$, 
$e^{\frac{2\pi i kf(s)}{n}}=e^{i\theta }$. 
\begin{ex}
Suppose that $f$ is $k$-constant and that,
for simplicity,  $\theta=0$. 
Then $f(s)=\frac{nr}{k}$ for some integer
$r$, and for all $s\in X$. For example let $n=8$. Then $f$ is $1$-constant if and only if
$f\equiv 0$; $f$ is
$2$-constant if and only if $f(X)\subset \{0,4\}$ and  $f$ is
$4$-constant if and only if $f(X)\subset \{0,2,4,6\}$. To say that $f$ is 
$3$-constant, $5$-constant or $7$-constant means that $f\equiv 0$. To say that $f$ is $6$-constant means that
$f(s)=\frac{4r}{3}$, which means that $f(s)\subset\{0,4\}$, for all
$s\in X$.
\end{ex}
%\subsection{Generalised Balanced Functions}
\begin{defn} Let $\rho^k$ be an irreducible (unitary) representation of $H$. Let 
$n=\dim \rho^k$ and suppose that $i\in \{1,\ldots,n\}$.
We say that $f:X\ra H$ is $\rho_i^k$-{\em balanced} 
if for all $r\in \{1,\ldots,n\}$  we have
$\sum_{g \in X}{\tau_{i,r}^k(f(g))}=0$. 
\end{defn}
As before, we can refer to $f$ being ``$\chi$-balanced'' in the case where $\chi$ is 
a linear representation of $H$. 

The trivial representation $\chi_0$ of $H$ is the 
map sending every element of $H$ to $1\in \CC$. Therefore
$f$ can  never be
$\chi_0$-balanced and we usually consider only $\chi$-balanced 
functions $f$ where $\chi$ is non-trivial.

Again, if $f:X\ra\ZZ_n$ then 
to say that $f$ is $k$-balanced is to say that $\sum_{s\in X} e^{\frac{2\pi ikf(s)}{n}}=0$. 
\begin{ex}\label{ex:bal}
Suppose that $X=H=\ZZ_n$, $k=1$ and $n=8$. One possiblility is that $f$ is surjective,
but this is not necessarily the case. 
For example $f$ could take four values of $1$ and four values of $5$.
In Figures \ref{fig:kbalanced1} and \ref{fig:kbalanced2} 
we illustrate these possibilites, showing each of the eighth roots of
unity labelled with the number of elements of $X$ mapping to it under
$\chi_1\circ f$. Two of the other possibilities are illustrated in 
Figure \ref{fig:kbalanced3}, where
$f$ takes
values $1$,$3$,$5$ and $7$ twice each, and in Figure \ref{fig:kbalanced4},
where
$f$ takes the values $2$ and $6$ once each and the values $1$ and $5$
three times each.
\begin{figure}
  %\begin{center}
    \psfrag{1}{$1$}
    \psfrag{4}{$4$}
    \psfrag{2}{$2$}
    \psfrag{3}{$3$}
    \subfigure[Surjective]
    {\label{fig:kbalanced1}
     \includegraphics[scale=.4]{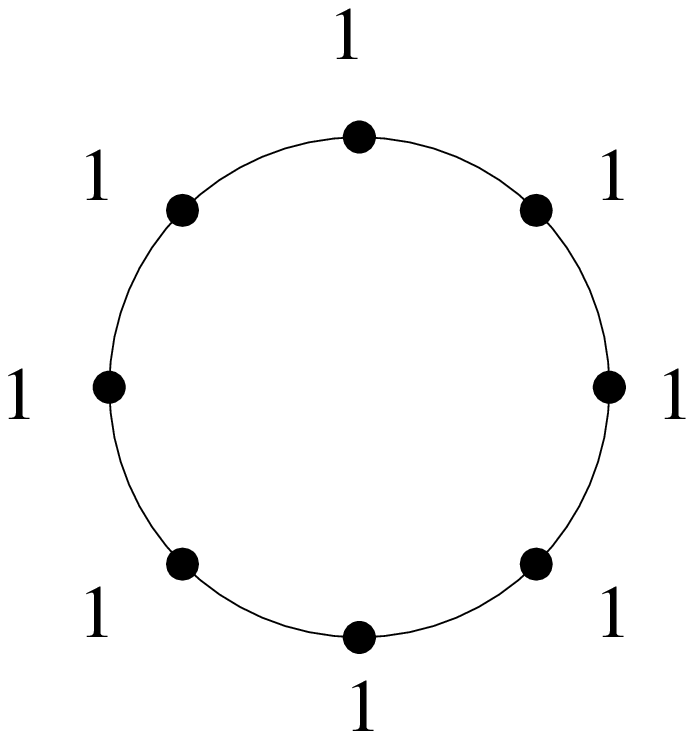}
     \hspace*{.03\textwidth}
    }
        \subfigure[$2$-valued]
    {\label{fig:kbalanced2}
      \psfrag{1}{\color{white}$1$}
      \includegraphics[scale=.4]{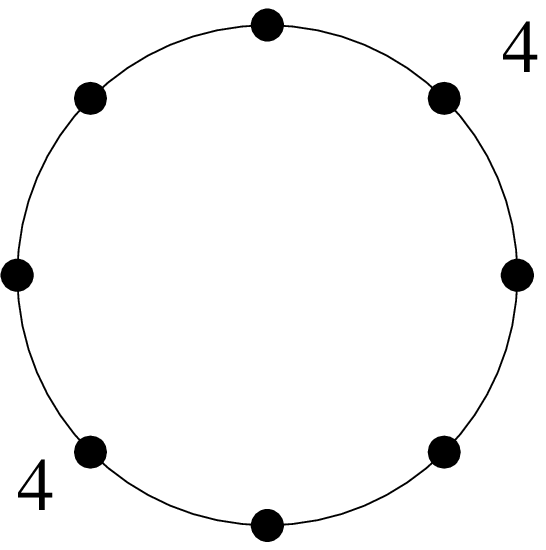}
      \hspace*{.03\textwidth}
    }
    %\hspace*{.05\textwidth}
    \subfigure[$4$-valued, evenly]
    {\label{fig:kbalanced3}
      \psfrag{1}{\color{white}$1$}
      \includegraphics[scale=.4]{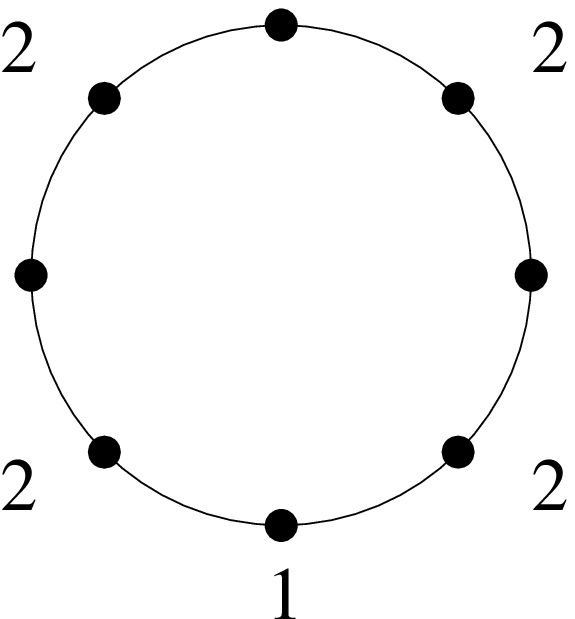}
      \hspace*{.03\textwidth}
    }
    %\hspace*{.05\textwidth}
    \subfigure[$4$-valued, unevenly]
    {\label{fig:kbalanced4}
      \psfrag{1}{\color{black}$1$}
      \includegraphics[scale=.4]{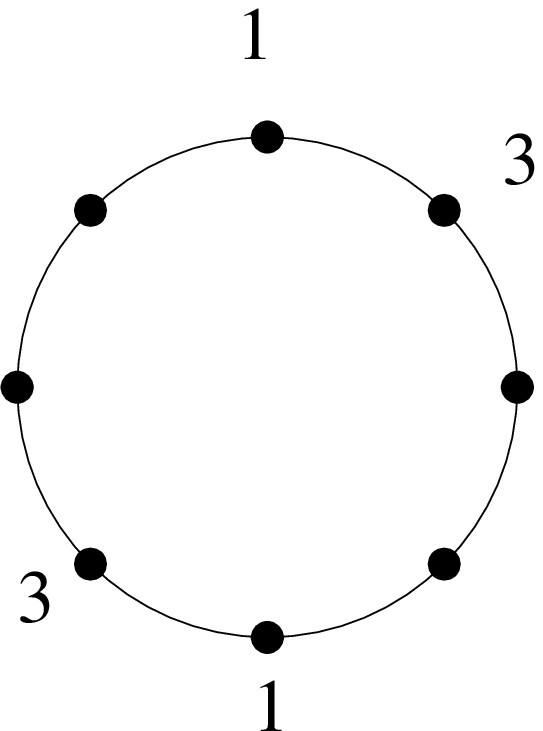}
    }
    \caption{$1$-balanced functions into $\ZZ_8$.}\label{fig:kbalanced}
  %\end{center}
\end{figure}
\end{ex}
%%%%%%%%%
\begin{comment}
\begin{figure}[!btph]
  \begin{center}
    \psfrag{1}{$1$}
    \includegraphics[scale=0.5]{s8.eps}
    \caption{\label{fig:kbalanced1} A $1$-balanced function which is surjective.}
  \end{center}
\begin{figure}[!btph]
\begin{center}
\includegraphics[width=2.0in]{kbalanced2.eps}
\caption{\label{fig:kbalanced2} A $1$-balanced function taking two values evenly.}
\end{center}
\end{figure}
\begin{figure}[!btph]
\begin{center}
\includegraphics[width=1.0in]{kbalanced3.eps}
\caption{\label{fig:kbalanced3} A $1$-balanced function taking four values evenly.}
\end{center}
\end{figure}
\begin{figure}[!btph]
\begin{center}
\includegraphics[width=1.5in]{kbalanced4.eps}
\caption{\label{fig:kbalanced4} A $1$-balanced function taking two values but not evenly.}
\end{center}
\end{figure}
\end{comment}

The definitions of $\rho^k_i$-constant and balanced functions are in a form
convenient for computation, as we'll see in Section
\ref{section:algorithm}. By contrast, the following characterisations of 
such functions,  
 in terms of the
integral group ring of $H$, emphasise their
structural
properties. Let $\ZZ H$ denote the integral group ring of $H$, $\ZZ
H=\oplus_{h\in H}\ZZ h$. 
If $T$ is a subset of $H$ then define $\ZZ T=\sum_{t\in T} \ZZ t$. 
As
usual, by an
$H$-module we mean a $\ZZ H$-module. Recall that if $\rho$ is a representation of
$H$ of dimension $n$ then $H$ acts on the right on $\CC^n$ by 
$$v\cdot h=v\rho(h),\textrm{ for } v\in \CC^n\textrm{ and } h\in H,$$
where we regard $v$ as a row-vector of length $n$ and $\rho(h)$ as an
$n\times n$ matrix over $\CC$. This action of $H$ extends by linearity
to an action of $\ZZ H$ on $\CC^n$; which is in this way a right
$H$-module. For $v\in \CC$ define the {\em annihilator} of $\langle
v \rangle$ 
(with respect to $\rho$) to be 
$$\ann (v)=\{r\in \ZZ H:v\cdot r =0\}.$$ Then $\ann(v)$ is a right
ideal of $\ZZ H$. We also define the {\em stabiliser}, in $H$, of an element
$v\in\CC^n$ to be
$$\stab_H(v)=\{h\in H:v\cdot h= v\}.$$

Since $H$ is finite we may assume that $H=\{h_1,\ldots
,h_d\}$, where $d=|H|$. Given $f:X\ra H$ define
$m_j=|f^{-1}(h_j)|$, $j=1,\ldots ,d$; so $m_j\ge 0$ and
$\sum^d_{j=1}m_j=|X|$.
 We call an element $r=\sum_{j=1}^d a_j h_j$ of $\ZZ H$ 
{\em  admissible}
if $a_j\ge 0$ and $\sum_{j=1}^d a_j=|X|$.
\begin{defn}\label{defn:r_f}
Given $f:X\ra H$ the element
$$r_f=\sum_{j=1}^d m_j h_j \in \ZZ H$$
is called the element of $\ZZ H$ {\em associated} to $f$.
\end{defn}
We denote the $i$th standard basis
element, the row-vector which is zero everywhere except the $i$th
coordinate
which is $1$, by $e_i$. 
Given an irreducible representation $\rho$ of $H$ we define 
$$\tau=\sqrt{\frac{\dim (\rho)}{|H|}} \rho.$$ 
This is consistent with the definitions of Section \ref{subsection:reps} since,
using the notation of that section, we have
$\tau_{i_h,j_h}=\tau_H^h$.
The first statement of the following Theorem is due
to S. Linton. 
\begin{theo}\label{theo:const_bal_gen}
Let $f:X\ra H$ be a map, let $\rho$ be an irreducible representation
of $H$, let $r_f$ be the element of $\ZZ H$ associated to $f$  and let
$S=\stab_H(e_i)$. Then
\begin{enumerate}[(i)]
\item $f$ is $\rho_i$-constant if and only if $r_f\in \ZZ T$, where
  $T$ is a 
  coset $T=Sh$ of $S$ in $H$, with $h\in \im(f)$, and 
\item $f$ is $\rho_i$-balanced if and only if $r_f\in \ann(e_i)$.
\end{enumerate}
\end{theo}
\begin{proof}
By  definition $f$ is $\rho_i$-constant if and only if there exists 
 $c=(c_1,\ldots ,c_r)\in \CC^n$ such that
$$(\tau_{i,1}(f(g)),\ldots, \tau_{i,n}(f(g))=c,$$
for all $g\in X$. The left hand side of the equality above is  
$e_i\tau(f(g))$, so $f$ is $\rho_i$-constant if and only if
$$e_i\cdot f(g)=e_i\rho(f(g))=c^\prime,$$
where $c^\prime =\sqrt{|\dim(\rho)|/|H|} c$, for all $g\in X$. 
Choose $h\in \im(f)$; so $e_i\cdot h=c^\prime$. If $h^\prime \in H$
then
$e_i\cdot h^\prime =c^\prime= e_i\cdot h$ if and only if $h^\prime
=sh$, for some $s\in S$. Thus $\{h^\prime\in H:e_i\cdot
h^\prime=c^\prime\}=T$, where $T=Sh$. It follows that $f$ is $\rho_i$-constant if and only if
$\im(f)\in T$; if and only if $r_f\in \ZZ T$.  As $h$ is an arbitrary element
of $\im(f)$ the first statement of the theorem now follows.

The function $f$ is $\rho_i$-balanced
if and only if 
$$0=\sum_{g\in X}\tau_{i,r}(f(g))=\sum_{j=1}^d m_j \tau_{i,r}(h_j),$$
for $r=1,\ldots ,n$. That is, if and only if 
$$0=\sum_{j=1}^d m_j  (\tau_{i,1}(h_j),\ldots ,\tau_{i,n}(h_j))=
e_i\sum_{j=1}^d m_j\tau (h_j).$$
Since $\tau$ and $\rho$ differ only by a constant this holds if and
only if 
$$0=e_i\sum_{j=1}^d m_j\rho (h_j)=e_i\cdot \sum_{j=1}^d m_j h_j,$$
that is if and only if $r_f\in \ann(e_i)$, as required.
\end{proof}

As is clear from the proof above $f$ is $\rho_i$-constant if and only
if $\im(f)\subset T$, where $T$ is an appropriate coset of
$\stab_H(e_i)$. Thus we may characterise $\rho_i$-constant functions
without
reference to the group ring. However there does not appear to be such
a simple characterisation of $\rho_i$-balanced functions, for which we
need to pass to the group ring. To compare the two we then need to
recast
the characterisation of $\rho_i$-constant in similar terms.

Note that if $X=\{x_1,\ldots ,x_n\}$ then, classically, we may compute
$f(x_j)$, for $j=1,\ldots ,n/2$, and find that $e_i\cdot f(x_j)=c$,
for all such $j$. If $f(x_{j+1})$ is such that $e_i\cdot f(x_{j+1})=c$
then $f$ is $\rho_i$-constant. However, if $e_i\cdot f(x_j)=-c$, for
$j=n/2+1, \ldots ,n$, then $f$ is $\rho_i$-balanced. Hence we can
distinguish, with certainty, between $\rho_i$-constant and
$\rho_i$-balanced functions,
using classical computation, only after making $n/2+1$ calls to the
oracle for $f$. Therefore a classical algorithm cannot solve this prolem
in polynomial time.
%%% copy a bit more of section below here
In Section \ref{section:algorithm} we show that for
the same purpose
a quantum algorithm requires only one call to the quantum oracle
for $f$. Thus, as in the case of the standard Deutsch--Jozsa
algorithm, quantum computation gives an improvement in speed
which seems impressive. However, as the number of admissible elements
of $\stab_H(e_i)$ is much smaller, in general, than the number of
admissible elements in $\ann(e_i)$,  using a classical algorithm
we can quickly distinguish between a  $\rho_i$-constant and
$\rho_i$-balanced functions, to within a bounded probability of error.
To be more exact: in Section \ref{section:cases} below we observe that the original
Deutsch--Jozsa algorithm may  be viewed as a  special case of our algorithm. 
In this case 
we can use a classical algorithm to 
determine whether $f$ is $\rho_i$-constant or $\rho_i$-balanced, with
probability of error less than $1/2$, in two calls to the oracle evaluating $f$
(see for example \cite{nc}): so the  problem lies in the complexity
class BPP. Hence in general if we accept bounded error computation
then our quantum algorithm only gives a constant-factor improvement over a classical
algorithm. However it should be emphasised the 
quantum algorithm gives and exact answer, so the more general problems described
here lie in complexity class EQP. 
\begin{comment}
Suppose now that $r=\sum_{j=1}^d a_j h_j$ is an admissible element of $\ZZ
H$. Partition $X$ into $d$ disjoint subsets $X_1,\ldots , X_d$ such 
that $|X_j|=a_j$. Define a function $g:X\ra H$ by $g(x)=h_j$ if and
only if $x\in X_j$. Then $r_g=r$. Therefore every admissible element
of $\ZZ H$ is associated to {\bf some number which is well-known (but
not to me)} functions $X\ra H$.
\end{comment} 

\subsection{Finite Abelian Groups}\label{sub:abelian}
In Section \ref{section:qft} we made use of a bijection $\beta:G\ra R$, where
$R$ is a set which indexes all the matrix entries of the unitarized irreducible representations
of $G$. In the general case there is no canonical choice of $\beta$.
In some cases, however it is clear which bijection to choose, and this lends extra structure to
the Fourier transform. One such case is that of an Abelian group $A$, where every irreducible 
representation is one-dimensional. In this case the
irreducible representations coincide with the characters of $A$ and 
form a group, denoted $\hat A$, of the same order as $A$. Suppose that $A=\ZZ_n$ 
is a cyclic group, under addition $\mod$ $n$ and let the characters of
$A$ be $\rho^k$, where
$\rho^k(a) =e^{2\pi i ak/n}$, $k=0,\ldots,n$. Then $k\mapsto \rho^k$ is an isomorphism
between $A$ and $\hat A$.
%We note also that if $G_1$ and $G_2$ are abelian and
%we have canonical bijections $\beta_1:G_1\ra X_1$ and $\beta_2:G_2\ra X_2$ then we have a 
%canonical bijection $\beta:G_1\times G_2\ra X_1\times X_2$ which we define to be
%$\beta=\beta_1\times\beta_2$. Then $\rho^{(g_1,g_2)}=\rho^{g_1}\tens\rho^{g_2}$.
%Since the representations are one-dimensional, this is equal to $\rho^{g_1}\rho^{g_2}$.

This generalises to the case where $A$ is an arbitrary finite Abelian group, say
$A=\oplus_{j=1}^k\ZZ_{n_j}$, of order $n=\prod n_j$, as follows.
Let $m=(m_1,\ldots,m_k)\in A$ and let $\rho^{j,i}$ be the $i$th
character of $\ZZ_{n_j}$, as above. 
Then  the map
$m=(m_1,\ldots,m_k)\mapsto \prod_{j=1}^k\rho^{j,m_j}=\rho^m_A$ is an isomorphism between
$A$ and $\hat A$. 
\begin{comment}
Now, given an integer $m$ with $0\le m<n$ we can write it uniquely in the form
\[ m=m_1+m_2n_1+m_3n_1n_2+\cdots+m_k\prod_{r=1}^{k-1}n_r,
\] 
with $0\le m_j<n_j$, $j=1,\ldots,k$.
%(This is reminiscent of the process of expressing an integer in a certain number base, but with a
%different base each time the exponent is increased).
So, for $0\le m<n$ and $g=(g_1,\ldots,g_k)\in X$,  we have
\begin{align*} \tau^m(g) & =
  \frac{1}{\sqrt{n}}\prod_{j=1}^k\rho^{j,m_j}(g_j)
=\frac{1}{\sqrt{n}}\prod_{j=1}^ke^{2\pi i g_jm_j/n_j}\\
%&=\frac{1}{\sqrt{n}}\prod_{j=1}^ke^{2\pi i g_jm_j/n_j}. \\
& =  \frac{1}{\sqrt{n}}exp\left(2\pi i\sum_{j=1}^k\frac{g_jm_j}{n_j}\right).
\end{align*}
In the particular case where $n_j=2$ for all $j$ we have
\begin{align*}
\tau^m(g) & =  \frac{1}{\sqrt{2^k}}(-1)^{\sum_{j=1}^kg_jm_j}\\
& =  \frac{1}{\sqrt{2^k}}(-1)^{g\cdot m}.
\end{align*}
\notes{AJD: where do we need this? Should it be in the section on QFT?}
\end{comment}
So, for fixed $m$ and all $a=(a_1,\ldots,a_k)\in A$,  we have
\begin{align} \tau_A^m(a) & =
  \frac{1}{\sqrt{n}}\prod_{j=1}^k\rho^{j,m_j}(a_j)
=\frac{1}{\sqrt{n}}\prod_{j=1}^ke^{2\pi i a_jm_j/n_j}\nonumber\\
%&=\frac{1}{\sqrt{n}}\prod_{j=1}^ke^{2\pi i g_jm_j/n_j}. \\
& =  \frac{1}{\sqrt{n}}\exp\left(2\pi
  i\sum_{j=1}^k\frac{a_jm_j}{n_j}\right).\label{al:tau}
\end{align}
Now set $k_j=n/n_j$, for $j=1,\ldots ,k$ and define 
$$\phi_m:A\ra \ZZ_n\textrm{ by }\phi_m(a)=\left(\sum_{j=1}^k
  a_jm_jk_j\right) \mod n.$$ 
Then $\phi_m$ is a well-defined map from $A$ to $\ZZ_n$ (which is a
homomorphism
but not, in general, an isomorphism). Define $f_m=\phi_m\circ f$, a
map
from $X$ to $\ZZ_n$. From \eqref{al:tau}
\begin{align*}
\tau_A^m(a) & = \frac{1}{\sqrt{n}}\exp\left(\frac{2\pi i}{n}
  \sum_{j=1}^k a_jm_jk_j\right) \\
& = \frac{1}{\sqrt{n}}\exp\left(\frac{2\pi i}{n}\phi_m(a)\right)\\
& = \tau^1_{\ZZ_n}(\phi_m(a)).
\end{align*}
Therefore $\tau_{\ZZ_n}^1\circ f_m=\tau^m_A\circ f$ and we have the
following lemma.
\begin{lemma}
In the notation above, $f$ is $m$-constant if and only if $f_m$ is
$1$-constant and $f$ is $m$-balanced if and only if $f_m$ is $1$-balanced.
\end{lemma}
\noindent
In the light of this lemma, if the codomain of $f$ is Abelian we may
always assume that it is cyclic.

We shall now analyse more carefully the condition that $f:X\ra H$ is
$k$-constant or $k$-balanced when $H$ is the finite cyclic group
$\ZZ_n$ and $0\le k <n$. Let $d=\gcd(k,n)$, and suppose that $k=ud$, $n=vd$. Let
$\langle v \rangle$ be the subgroup of $\ZZ_n$ generated by $v$. Then
$\ZZ_v\cong \ZZ_n/ \langle v\rangle$, and there is a canonical
homomorphism $\pi:\ZZ_n\ra \ZZ_v$. Let $\bar f=\pi\circ f$, so $\bar
f(a)=f(a) \mod v$, for $a\in\ZZ_n$.
\begin{prop}\label{prop:kconst}  
In the notation above the following are equivalent.
\begin{enumerate}[(i)]
\item\label{lemma:v1} $f$ is $k$-constant.
\item\label{lemma:v2}  $\bar f$ is $u$-constant.
\item\label{lemma:v3} $\bar f$ is constant.
\end{enumerate}
\end{prop}
\begin{proof} 
As observed following Definition \ref{defn:const}, $f$ is $k$-constant
if and only if there exists a constant $\theta\in\RR$ such that 
$e^{2\pi i kf(s)/n}= e^{2\pi i\theta/n}$, for all $s\in X$. This is so if and
only if $kf(s)\equiv \theta \mod n$; if and only if $uf(s)\equiv
(\theta/d) \mod v$. As $f(s)\equiv \bar f(s)\mod v$, this shows that \eqref{lemma:v1} and
\eqref{lemma:v2} are equivalent.
Now $\bar f$ is $u$-constant if and only if  $uf(s)\equiv
\theta \mod v$, for some $\theta$, if and only if $f(s)\equiv
u^{-1}\theta \mod v$, as $u$ and $v$ are coprime. Thus
\eqref{lemma:v2} and \eqref{lemma:v3} are equivalent.
\end{proof}
%%%%%%%%%%%%%%%%%%%%%%%
\begin{corollary}\label{corollary:const}
$f$ is $k$-constant if and only if $f(X)$ is contained within some
coset of $\langle v\rangle$.
\end{corollary}
\begin{proof} 
$f$ is $k$-constant if and only if $\bar f$ is constant, from Proposition
\ref{prop:kconst}\eqref{lemma:v3},
and the result follows.
\end{proof}
\begin{corollary}\label{corollary:primeconstant} 
If $p$ is a prime number, then for all $k\in\{1,\ldots,p-1\}$, a function
$f:\ZZ_p\ra\ZZ_p$ is $k$-constant if and only if it is constant.
\end{corollary}
\begin{proof} This follows directly from the equivalence of Proposition
\ref{prop:kconst}\eqref{lemma:v2} and \eqref{lemma:v3}.
\end{proof}

\begin{prop} \label{prop:kbal}
In the notation above the following are equivalent.
\begin{enumerate}[(i)]
\item\label{prop:k1} $f$ is $k$-balanced 
\item\label{prop:k2} $\bar{f}$ is $u$-balanced.
\item\label{prop:k3} $\bar f$ is $1$-balanced.
\end{enumerate}
\end{prop}
\begin{proof}
\eqref{prop:k1} and \eqref{prop:k2} are equivalent because
\[ \sum_{s\in X} e^{2\pi i kf(s)/n}=\sum_{s\in X} e^{2\pi iuf(s)/v}
=\sum_{s\in X} e^{2\pi i u\bar{f}(s)/v}.
\]
To see the equivalence of \eqref{prop:k2} and \eqref{prop:k3} note
that, because gcd$(u,v)=1$, \[ \{0,\ldots,v-1\}=_{\ZZ_n}\{0,u,2u,\ldots,(v-1)u\},
\]
i.e. $0,u,\ldots,(v-1)u$ is a complete set of residues for $\ZZ_v$. 
Therefore
$\sum_{s\in X}e^{2\pi i uf(s)/v}=\sum_{s\in X} e^{2\pi i f(s)/v}$.
\end{proof}

Given $n$ and $k$ as above,
replacing the function $f:X\ra \ZZ_n$ with the function $\bar f:X\ra
\ZZ_v$, it follows from Propositions \ref{prop:kconst} and
\ref{prop:kbal}, that we reduce the problem of distinguishing between
$k$-constant and $k$-balanced to that of distinguishing between
constant
and $1$-balanced. Therefore we now restrict to functions $f:X\ra
\ZZ_n$ which are either constant or $1$-balanced. 
 
Corollary \ref{corollary:const} gives a characterisation of 
$k$-constant functions in terms of the subgroup $\langle v \rangle$ of
$\ZZ_n$
but, despite the similarities between Propositions \ref{prop:kconst}
and
\ref{prop:kbal}, we have no analagous characterisation of $k$-balanced
functions. In order to find such a characterisation it is convenient
to recast Definition \ref{defn:r_f} in terms of polynomials over $\ZZ$, since
in this special case we obtain a polynomial of one variable.
As before, given $f:X\ra \ZZ_n$ we may
define the integer 
$p_t=|f^{-1}(t)|$, for $t=0,\ldots,n-1$, and now define the polynomial 
\[P_f(x)=\sum_{t=0}^{n-1}p_tx^t.\] 
(Regarding $x$ as the generator of $\ZZ_n$ we may identify $P_f$ with the element $r_f$ of the integral group
ring of $\ZZ_n$.)
Observe that 
\begin{enumerate}[(a)]
\item\label{it:a} the degree 
of $P_f$ is  at most $n-1$, 
\item\label{it:b} all the coefficients $p_t$ of are
non-negative, and 
\item\label{it:c} $\sum_{t=0}^{n-1}p_t=|X|$.
\end{enumerate}
Let $\omega=e^{2\pi i/n}$, then $f$ is $1$-balanced 
if and only if 
\[0=\sum_{s\in X} \omega^{f(s)}=\sum_{t=0}^{n-1}p_t \omega^t=P_f(w).\]
The minumum polynomial of $\omega$ over $\QQ$ is $\Phi_n$, the $n$th
cyclotomic polynomial (see Section \ref{section:appendix} for further details).
%; which has degree $\phi(n)$. 
Therefore $f$ is $1$-balanced if and only if
$\Phi_n|P_f$.
On the other hand $f$ is constant if and only if $P_f$ is a monomial
(i.e. has the form $p_tx^t$, for some $t$).
Conversely, given a polynomial $P$ satisfying 
\eqref{it:a}, \eqref{it:b} and \eqref{it:c} 
we
may define a function $f:X\ra \ZZ_n$, by choosing a partition of
$X$ into (at most) $n$ subsets $X_0,\ldots ,X_{n-1}$, such that $X_i$ has
size $p_i$, and defining $f(g)=t$, if and only if $g\in X_t$. Then 
$f$ is constant if and only if $P$ is monomial and is $1$-balanced if
and
only if $P$ is divisible by $\Phi_n$. If we regard the oracle for $f$
as an oracle which determines the polynomial $P_f$ then the promise
that $f$ is constant or $1$-balanced is equivalent to the promise that
$P_f$ is monomial of divisible by $\Phi_n$. The problem of
distinguishing
between constant or $1$-balanced functions is therefore equivalent to
the problem of distinguishing between (hidden) polynomials which are
either monomial or divisible by $\Phi_n$. 

\begin{ex}\label{ex:balpol}
Consider the functions $f:\ZZ_n\ra \ZZ_n$ of Example \ref{ex:bal},
as illustrated in Figure \ref{fig:kbalanced}.
For the function $f$ of Figure \ref{fig:kbalanced1}
$P_f=1+x+x^2+x^3+x^4+x^5+x^7+x^8$. 
For $f$ in Figures \ref{fig:kbalanced2}, \ref{fig:kbalanced3} and
\ref{fig:kbalanced4} we have $P_f=4x(1+x^4)$, $P_f=2x(1+x^2+x^4+x^6)$ and 
$P_f=x(3+x+3x^4+x^5)$, respectively. 
\end{ex}

We are now in a position to apply Theorem \ref{theo:dephi}, of Section
\ref{section:appendix}, and the following definition to characterise $1$-balanced
functions into $\ZZ_n$ for sufficiently simple $n$.
\begin{defn}\label{defn:even}
Let $X$ and  $Y$ be sets, $S$ a subset of $Y$ and $f:X\ra Y$ a function from $X$ to
$Y$. Then $S$ is {\em evenly covered} by $f$ if there exists $m\in\ZZ$
such that $|f^{-1}(s)|=m$, for all $s\in S$.
\end{defn}
In keeping with the terminology of \cite{hoyer2}, if $Y$ is evenly covered by $f$ 
we shall say that $f$ is {\em perfectly balanced}.
\begin{theo}\label{theo:even}
Let $n$ be a positive integer and let $p$ and $q$ be distinct primes
such that $n=p^\alpha q^\beta$, where $\alpha$ and $\beta $ are
integers,
$\alpha>0$ and $\beta\ge 0$. Let $X$ be a finite set and $f:X\ra \ZZ_n$
a function. Define $K_p$ to be the subgroup of $\ZZ_n$ generated by
$n/p$ and, if $\beta >0$, define $K_q$ to be the subgroup
generated by $n/q$.
\begin{enumerate}[(i)]
\item\label{item:even1} If $\beta =0$ then
$f$ is  $1$-balanced if and only if every coset of 
$K_p$ is evenly covered by $f$.
\item\label{item:even2} If $\beta>0$ then $f$ is  $1$-balanced if and only
if there exists a partition of $X$ into disjoint subsets $X_p$ and
$X_q$ such that every coset of $K_p$ is evenly covered by $f|_{X_p}$
and every coset of $K_q$ is evenly covered by $f|_{X_q}$.  
\end{enumerate}
\end{theo}
\begin{remark}
The obvious generalisation of this theorem to integers with $3$ or
more
prime factors does not hold, as shown by
Example \ref{ex:balnode} below. The best we have been able to do is
Proposition \ref{prop:genbal}.
\end{remark}
\begin{proof}
From the discussion above the function $f$ is $1$-balanced if and only
if $P_f$ is divisible by $\Phi_n$. Consider first the case $\beta=0$.
From Theorem \ref{theo:dephi}, we have  
$P_f(x)=s(x)\Phi_p(x^{n/p}),$ where $s\in \ZZ[x]$ and the
coefficients of $s$ are all non-negative. As deg$(P_f)\le n-1$ and deg$(\Phi_p)=p-1$ it
follows that deg$(s)\le n/p -1$. Let $s(x)=u_0+u_1(x)+\cdots
+u_{n/p-1}x^{n/p-1}$. Fix $t\in\ZZ$ with $0\le t<n-1$. Since $\Phi_p(x)=1+x+\cdots +x^{p-1}$
the coefficient $p_t$ of $x^t$ in $P_f$ is $u_j$, where $j$ is the unique
integer such that $j\equiv t \mod n/p$ and $0\le j<n/p$. Therefore
the coefficient $p_t$ equals the coefficient $p_r$, for all $r$ such
that $r\equiv t\mod n/p$. Thus, if $0\le t<n/p$, we have 
$p_t=p_r$, for $r=t,n/p+t, \ldots, (p-1)n/p +t$.
As $p_t=|f^{-1}(t)|$ it follows that the
coset $t+K_p$ is evenly covered by $f$. The converse follows easily,
by reversing this argument.

Now suppose that $\beta>0$. This time Theorem \ref{theo:dephi} implies that
$f$ is $1$-balanced if and only if 
$P_f(x)=s_1(x)\Phi_p(x^{n/p})+s_2(x)\Phi_q(x^{n/q})$, where 
$s_i\in\ZZ[x]$ and the coefficients of $s_i$ are all non-negative.
Let $A(x)=s_1(x)\Phi_p(x^{n/p})$ and $B(x)=s_2(x)\Phi_q(x^{n/q})$, and
suppose that $A(x)=a_0+a_1 x+\cdots + a_{n-1}a^{n-1}$ and 
$B(x)=b_0+b_1x+\cdots +b_{n-1}x^{n-1}$. As in the case $\beta =0$ the
coefficients $a_r$ and $a_t$ are equal for all $r,t$ such that 
$0\le r,t<n$ and $r\equiv t \mod n/p$. A similar statement, involving
$q$ instead of $p$, holds for the coefficients of $B$. For fixed $t$
we
have $|f^{-1}(t)|=a_i+b_j$, where $i\equiv t \mod n/p$ and $j\equiv t
\mod n/q$. Hence we may partition $f^{-1}(t)$ into disjoint (possibly
empty) subsets 
$X_{p,t}$ and $X_{q,t}$ such that  $|X_{p,t}|=a_i$ and $X_{q,t}=b_j$. 
Now $t\equiv r \mod n/p$ implies $a_r=a_t$ so also
$|X_{p,t}|=|X_{p,r}|$. Setting $X_p=\cup_{t=0}^{n-1}X_{p,t}$ we see
that
$f|_{X_p}$ covers $t+K_p$ evenly, for $t=0,\ldots ,p-1$. Similarly, if
$X_q=\cup_{t=0}^{n-1}X_{q,t}$
then $f|_{X_q}$ covers $t+K_q$ evenly, for $t=0,\ldots ,q-1$. As
$X=X_p\cup X_q$ and $X_p\cap X_q=\emptyset$, this completes the proof
of the theorem.
\end{proof}
\begin{comment}
\notes{AJD: If $G$ is a set, rather than a group, then we can continue
  to decompose the functions from $f|_{G_{p,t}}$ to cosets of $K_p$ 
  in the same way; regarding each
  coset as a copy of $\ZZ_{n/p}$; if it turns out
  they're also $1$-balanced.
  This might be a way of doing a
  Bernstein-Vazirani.}
\end{comment}
\begin{prop}\label{prop:genbal}
Let $n$ be a positive integer with prime factorisation
$p_1^{\alpha_1}\cdots p_k^{\alpha_k}$. 
Let $K_{p_i}$ be the subgroup of $\ZZ_n$ generated by $n/p_i$.
Let $f:X\ra \ZZ_n$ be 
a function with associated polynomial $P_f$ such that
\[P_f(x)=\sum_{i=1}^k s_i(x)\Phi_{p_i}(x^{n/p_i}),\]
where $s_i\in \ZZ[x]$ and the coefficients of $s_i$ are all non-negative.
Then  $f$ is $1$-balanced and there exists a partition of $X$ into
disjoint subsets $X_1,\ldots ,X_k$ such that $f|_{X_i}$ evenly covers
the
cosets of $K_{p_i},$ $i=1,\ldots ,k$. Moreover,
setting $N_i$ equal to 
the sum of the coefficients of $s_i$
we have 
$|X_i|=nN_i/p_i$.
\end{prop}
The proof of Proposition \ref{prop:genbal} is similar to (the appropriate part of) the proof of Theorem
\ref{theo:even} and we leave the details to the reader.
\begin{ex}
Consider the polynomials of Example \ref{ex:balpol} corresponding to the
functions of Example \ref{ex:bal} and Figure \ref{fig:kbalanced}. Here
$K_p=K_2=\langle 4 \rangle=\{1,4\}$.
For Figure \ref{fig:kbalanced1} we have
$P_f=1+x+x^2+x^3+x^4+x^5+x^7+x^8=(1+x+x^2+x^3)\Phi_2(x^4)$. In this
case every coset of $K_2$ is covered evenly by one
element of $X$. 
Corresponding to Figure \ref{fig:kbalanced2},
$P_f=4x(1+x^4)=4x\Phi_2(x^4)$. Here $1+K_2$ is evenly covered by $4$ elements
and all other cosets are covered by $0$ elements.
Figure
\ref{fig:kbalanced3} gives
$P_f=2x(1+x^2+x^4+x^6)=2x(1+x)\Phi_2(x^4)$. This time $K_2$ and
$3+K_2$ are covered
by $0$ elements and $1+K_2$ and $2+K_2$ by $2$ elements.
With Figure
\ref{fig:kbalanced4} we have 
$P_f=x(3+x+3x^4+x^5)=x(3+x)\Phi_2(x^4)$; the coset $1+K_2$ is covered
by $3$ elements, the coset $2+K_2$ is covered by $1$ element and both
other cosets by $0$ elements.
\end{ex}
\begin{ex}
Let $n=15$ and $f$ be a function $\ZZ_{45}\ra \ZZ_{15}$.
In this case $K_3=\langle 5 \rangle$ and $K_5=\langle 3 \rangle$.
 If 
$P_f=(4+2x+x^2+3x^4)\Phi_3(x^5)+(2+x^2)\Phi_5(x^3)$ then $f$ is
$1$-balanced. 
We can partition $X$ into subsets $X_3$ of size $30$ and 
$X_5$ of size $15$ such that $f|_{X_3}$ covers $K_3$ evenly with $4$
elements,
$1+K_3$ with $2$ elements, $2+K_3$ with $1$ element $3+K_3$ with $0$
elements and $4+K_3$ with $3$ elements. Similaraly $f|_{X_5}$ covers
cosets $t+K_5$, for $t=0,1,2$, evenly with $2$, $0$ and $1$ elements,
respectively.
\end{ex}
\begin{ex}\label{ex:balnode}
We are grateful to C. Smyth for pointing this example out to us.
Let $n=105$,  $\omega
=\exp{2\pi i/105}$, $\zeta = \omega^7$ and $\eta=\omega^{15}$, so
$\zeta^{15}=\eta^7=1$. The minimum polynomial of $\zeta$ over $\QQ$ is 
$\Phi_{15}(x)=1-x+x^3-x^4+x^5-x^7+x^8$, so we have
$1+\zeta^3+\zeta^5 +\zeta^8=\zeta+\zeta^4+\zeta^7$. The minimum polynomial
of $\eta$ is $\Phi_7(x)$ so we have $1+\eta+\eta^2+\cdots
+\eta^6=0$. Therefore
\[(\zeta+\zeta^4+\zeta^7)(\eta+\eta^2+\eta^3+\eta^4+\eta^5+\eta^6) +
(1+\zeta^3+\zeta^5 +\zeta^8)=0.\]
Writing this out as a polynomial in $\omega$ we obtain
$P=\sum_{t=0}^{104} p_t\omega^t=0$, where $p_t=1$, for $t= 0$, $4$, $13$,
$19 $,
$21 $,
$22 $,
$34 $,
$35 $,
$37 $,
$43 $,
$52 $,
$56 $,
$58 $,
$64 $,
$67 $,
$73 $,
$79 $,
$82$,
$88$,
$94$,
$97$ and $103$,
and $p_t=0$ otherwise.
Let $f$ be a function $\ZZ_{105}\ra \ZZ_{105}$ such
that $P_f=P$. Then $f$ is $1$-balanced, as $P(\omega)=0$. 
Any straightforward analogue of
Theorem \ref{theo:even}
would (at the least) assert that there is a subset $S$ of $\ZZ_{105}$ and a subgroup $K$ of
$\ZZ_{105}$, such that the
restriction of $f$ to $S$ covers every coset of $K$
evenly.
Since
$p_0=1$, this would imply that $f|_S$ covers $K$ evenly.
Thus $f$ should map one element of $\ZZ_{105}$ to each element  of
$K$.
Hence $p_t$ should  be equal to $1$ for $t$ equal to some
divisor 
of $105$ and all its multiples. This is clearly not the case so no
such generalisation of 
Theorem \ref{theo:even} 
exists.
\end{ex}
\begin{corollary}\label{corollary:primebalanced}
If $p$ is a prime number then a function $f:\ZZ_p\ra\ZZ_p$ is $1$-balanced 
if and only if it is  a bijection.
\end{corollary}
\subsection{Non-Abelian Examples.}\label{sub:nonabelian}
The following examples involve the symmetric groups $S_n$ and the alternating
group $A_4$. It follows from the results of \cite{mrr}
(see the end of Section \ref{section:qft}) that 
there are efficient implementations of the quantum Fourier transform for these groups.
Therefore efficient quantum algorithms for the problems of these examples may be
constructed.
\begin{ex}
Consider the simplest possible non-Abelian finite group, $S_3$, considered as a dihedral group 
and generated by a rotation $r$ and a reflection $s$.
The irreducible representations
of $S_3$ are $\rho^1$, the trivial representation, $\rho^2$, the alternating representation,
and $\rho^3$, the two-dimensional representation. The corresponding Fourier coefficients are 
given in the following table.
\begin{center}
\begin{tabular}{l|llllll}
& $1$ & $r$ & $r^2$ & $s$ & $t=r^2s$ & $u=rs$ \\ \hline
$\tau_{1,1}^1$ & $1/\sqrt{6}$ & $1/\sqrt{6}$ & $1/\sqrt{6}$ & $1/\sqrt{6}$ & 
$1/\sqrt{6}$ & $1/\sqrt{6}$ \\
$\tau_{1,1}^2$ & $1/\sqrt{6}$ & $1/\sqrt{6}$ & $1/\sqrt{6}$ & $-1/\sqrt{6}$ & 
$-1/\sqrt{6}$ & $-1/\sqrt{6}$ \\  
$\tau_{1,1}^3$ & $1/\sqrt{3}$ & $e^{2\pi i/3}/\sqrt{3}$ & $e^{-2\pi i/3}/\sqrt{3}$ & $0$ & 
$0$ & $0$ \\
$\tau_{1,2}^3$ & $0$ & $0$ & $0$ & $1/\sqrt{3}$ & $e^{-2\pi i/3}/\sqrt{3}$ 
& $e^{2\pi i/3}/\sqrt{3}$ \\
$\tau_{2,1}^3$ & $0$ & $0$ & $0$ & $1/\sqrt{3}$ & $e^{2\pi i/3}/\sqrt{3}$ 
& $e^{-2\pi i/3}/\sqrt{3}$ \\
$\tau_{2,2}^3$ & $1/\sqrt{3}$ & $e^{-2\pi i/3}/\sqrt{3}$ & $e^{2\pi i/3}/\sqrt{3}$ 
& $0$ & $0$ & $0$ \\
\end{tabular}
\end{center}
\begin{enumerate}
\item
First consider the alternating representation $\rho^2$, which is linear.
To say that a function $f:X\ra S_3$ is $\rho^2$-constant means that
the image of $f$ is contained in 
$\langle r \rangle$ or its coset $\langle r \rangle s$. To say that $f$ is 
$\rho^2$-balanced means that $|f^{-1}(\langle r \rangle)|=|f^{-1}(\langle r \rangle s)|$.
\item
Now consider the $2$-dimensional representation $\rho^3$. To say that $f:X\ra S_3$ is
$\rho_1^3$-constant means that for $i=1$ and $2$ there is a constant $c_i\in\CC$ such that for
all $g\in X$, $\tau_{1,i}^3(f(g))=c_i$. For $i=1$ or $2$, the table above
shows that $f$ has to be constant. 
Since one coset of $\langle r \rangle$ always maps to zero under a matrix coefficient of 
$\rho^3$, the meaning of $\rho_1^3$-balanced is that
\[\sum_{g\in f^{-1}(\langle r \rangle)}\tau_{1,1}^3(f(g))=0 \mbox{ and }
\sum_{g\in f^{-1}(\langle r \rangle s)}\tau_{1,2}^3(f(g))=0.\] 
In other words, setting $m_j=|f^{-1}(r^j)|$ and $n_j=|f^{-1}(r^js)|$, $j=0,1,2$,
\[\sum_{j=0}^2 m_j e^{2\pi i j/3}=0\mbox{ and }
\sum_{j=0}^2 n_je^{2\pi i j/3}=0.\]

Setting $P(x)=\sum_{j=0}^2 m_j x^j$ and $Q(x)=\sum_{j=0}^2 n_j x^j$ it
follows that $f$ is $\rho^3_1$-balanced if and only if $\Phi_3|P$ and
$\Phi_3|Q$. Let $X_1=f^{-1}(\langle r\rangle)$ and $X_2=f^{-1}(\langle
r\rangle s$; so $X$ is the disjoint union of $X_1$ and $X_2$, and set
$f_i=f|_{X_i}$. Then, as in Section \ref{sub:abelian}, it follows that
$f$ is
$\rho^3_1$-balanced if and only if $\langle r\rangle$ is evenly
covered by $f_1$ and $\langle r\rangle s$ is evenly covered by $f_2$.

In this case (in the terminology of Theorem \ref{theo:const_bal_gen}) 
$\ann(e_1)=\ann(e_2)$ the ideal of $\ZZ S_3$ generated by the 
element $1+r+r^2$. Hence $f$ is $\rho^3_1$-balanced if and only if it 
is $\rho^3_2$-balanced.

\end{enumerate}
\end{ex}
\begin{ex}
Let $S_m$ be the symmetric group on $m$ objects and  
let $A_m$ denote its alternating subgroup of index $2$. 
Let $\chi$ be the alternating character of $S_m$: that is $\chi$ is the
linear character of $S_m$ given by $\chi(h)=1$ if $h\in A_m$ and
$\chi(h)=-1$ otherwise. Let $f:X\ra S_m$ be a function and assume that
 we are 
promised that either (a) $\Ima(f)\subset A_m$ or $\Ima(f)\subset S_m-A_m$
or (b) $|f^{-1}(A_m)|=|f^{-1}(S_m-A_m)|$. Then $f$ is 
$\chi$-constant in case (a) and $\chi$-balanced in case (b). 
%Hence 
%we can distinguish between (a) and (b) with certainty and using a single quantum query.
\end{ex}
\begin{ex}

The alternating group $A_4$ may be regarded as 
the orientation-preserving group of symmetries of a regular
tetrahedron, whose $1$-skeleton is embedded in $\RR^3$ 
as diagonals of faces of a cube with vertices
$(\pm 1,\pm 1,\pm 1)$. This gives rise to a $3$-dimensional 
unitary irreducible representation $\rho$ of $A_4$
generated by matrices 
\[ N=\left( \begin{array}{rrr} 0&1&0\\0&0&1\\1&0&0 \end{array} \right) \mbox{ and }
   R=\left( \begin{array}{rrr} -1&0&0\\0&-1&0\\0&0&1 \end{array} \right).
\]
The elements of $A_4$ may then be listed as
$$I,N,N^2, R, RN, RN^2, NR, NRN, NRN^2, N^2R, N^2RN, N^2RN^2.$$ Then
$\stab_{A_4}(e_1)=\{I,N^2RN\}$. Therefore
a function $f:X\ra A_4$ is $\rho_1$-constant if and only if $f(X)$ is contained in one of
the cosets \[
\{I,N^{2}RN\},\{N,N^2RN^{2}\},\{N^2,N^2R\},\{R, NRN^2\},
\{RN,NR\},\{RN^2,NRN\}.\]

Calculation of $e_1M$, for each $M\in A_4$ in turn shows that
$\ann(e_1)$ is the subset of $\ZZ A_4$ consisting of elements
$\sum_{M\in A_4}m_M M$ such that
\begin{align*}
m_I-m_R+m_{N^2RN}-m_{NRN^2}&=0,\\
m_N-m_{RN}+m_{N^2RN^2}-m_{NR}&=0 \textrm{ and}\\
m_{N^2}-m_{RN^2}+m_{N^2R}-m_{NRN}&=0.
\end{align*}
Therefore $f$ is 
is $\rho_1$-balanced 
if and only if $r_f$ has such a form.
We may also characterise $\rho_i$-constant and
$\rho_i$-balanced 
functions in this way, for $i=2,3$ and the results are very similar.

\end{ex}
%I plan to re-interpret this in terms of formal language theory as I did with 
\section{The Deutsch--Jozsa--H\o yer  Algorithm with Generalised Promises}
\begin{figure}[!btph]
\begin{center}
\psfrag{x}{}
\psfrag{y}{}
\psfrag{uf}{$U_f$}
\psfrag{mq}{qu$X$}
\psfrag{kq}{qu$H$}
\psfrag{k0}{$\ket{\idgp_X}$}
\psfrag{kn}{$\ket{h_0}$}
\psfrag{yf}{}
\psfrag{W}{${\cal F}_X^{\phantom{\dagger}}$}
\psfrag{w2}{${\cal F}_H^{\phantom{\dagger}}$}
\psfrag{w3}{${\cal F}_{X}^{\dagger}$}
\psfrag{w4}{${\cal F}_H^{\dagger}$}
\includegraphics[width=3.0in]{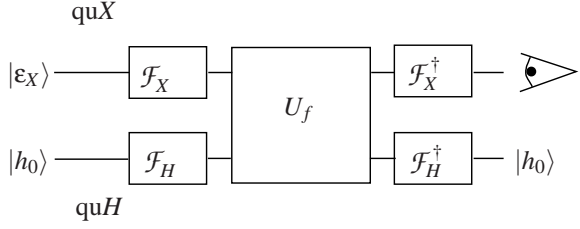}
\caption{\label{fig:groupcircuit} The quantum circuit for the Deutsch--Jozsa--H\o yer algorithm.}
\end{center}
\end{figure}

%the Z_2^n case. Maybe it will have some relevance to the word problem ...
\label{section:algorithm}
In this section we assume that we have a finite set $X$ and a finite
group $H$ and a map $f:X\ra H$, 
and we work with an oracle $U_f$ as in Section~\ref{section:oracles}.
We use the notation of Section \ref{subsection:reps} for
representations of the group $H$. In particular
let $H$ have irreducible unitary representations
$\rho^1,\ldots,\rho^R$ so we have 
\begin{equation*}%\label{eq:Htau}
\tau_{i,j}^k=\sqrt{\frac{\dim \rho^k}{|H|}}\cdot \rho^k_{i,j},
\end{equation*}
for $1\le k\le R$ and $1\le i,j\le \dim \rho^k$.
Then, since $\{\tau_h\}_{h\in H}$ is an orthonormal basis of $L^2(H)$,
we have 
\begin{equation}\label{eqn:tauorthog}
\langle \tau_{i,j}^k,\tau_{r,s}^t\rangle = \delta_{i,r}\delta_{j,s}\delta_{k,t}.
\end{equation}
\begin{lemma}\label{lemma:matrixmult}
Let $X$ be a finite set and $H$ be a finite group and $f:X\ra H$ be a map. Then, for
fixed
$i,j,k,r,s,t$, we have
\[ \sum_{h\in H}\tau_{i,j}^k(h)\overline{\tau_{r,s}^t(f(g)h)}=
\overline{\tau_{r,i}^k(f(g))}\delta_{j,s}\delta_{k,t}.
\]
\end{lemma}
\begin{proof}
Let $n=\dim \rho^t$. Using the formula for matrix multiplication,
\begin{align*}
\sum_{h\in H}\tau_{i,j}^k(h)\overline{\tau_{r,s}^t(f(g)h)}
& =  \sum_{h\in H}\tau_{i,j}^k(h)\sum_{q=1}^n\overline{\tau_{r,q}^t(f(g))\tau_{q,s}^t(h)} \\
& =  \sum_{q=1}^n \overline{\tau_{r,q}^t(f(g))}\sum_{h\in H}\tau_{i,j}^k(h)
\overline{\tau_{q,s}^t(h)} \\
& =  \sum_{q=1}^n\overline{\tau_{r,q}^t(f(g))}\langle\tau_{i,j}^k,\tau_{q,s}^t\rangle \\
& =  \sum_{q=1}^n\overline{\tau_{r,q}^t(f(g))}\delta_{i,q}\delta_{j,s}\delta_{k,t}
\mbox{ (by~(\ref{eqn:tauorthog})) } \\
& =  \overline{\tau_{r,i}^k(f(g))}\delta_{j,s}\delta_{k,t} \mbox{ as required.}
\end{align*}
\end{proof}
We shall use the
circuit in Figure~\ref{fig:groupcircuit}, which was introduced in \cite{hoyer2}, where it was used to
distinguish between perfectly balanced and constant functions. 
In order to apply the quantum Fourier transform to the $X$ register we
assume that $X=\ZZ_n$, where $n=|X|$. By $\idgp_X$ we mean the element of $X$ which corresponds
to $1_{\ZZ_n}$.
Our main result is Theorem~\ref{theo:main}, where it is shown that the range of promises that the algorithm can deal with extends 
beyond perfectly balanced and constant. 

If we omit the ${\cal F}_H^{\phantom{\dagger}}$ and ${\cal F}_H^{\dagger}$ gates in Figure~\ref{fig:groupcircuit}
and $h_0$ is set to be the identity, then we obtain the non-Abelian analogue of the circuit used in Shor's algorithm.
This has been proposed as a quantum algorithm for the hidden subgroup problem. 
We note, however, that it has been shown
in \cite{hrt-s} that while a polynomial number of Fourier samples will reconstruct 
a {\em normal} hidden subgroup,
the circuit fails to solve the hidden subgroup problem in $S_n$ for general subgroups, 
even in a very restricted 
situation (see also \cite{gsvv}, where the latter of these results was obtained independently).

When working with query complexity, only the number of calls to the oracle is relevant, and we do not discuss here the efficient
implementation of the Fourier transform on a finite group any further here 
(but see the end of 
Section \ref{section:qft}). 
\begin{theo}[General Form of the Deutsch--Jozsa--H\o yer Algorithm]\label{theo:main} 
Let $f:X\ra H$ be a function from the  finite set $X$ to the finite  group $G$  
and let $\rho^k$ be a non-trivial irreducible
representation of $H$. Let $n=\dim \rho^k$.
Suppose that we are promised that for some $i\in\{1,\ldots,n\}$, $f$ is either 
$\rho_i^k$-constant or $\rho_i^k$-balanced.
Then there exists a quantum algorithm which distinguishes between these two possibilities
with certainty, using a single quantum query.
\end{theo}
\begin{proof}
Let $j\in \{1,\ldots,n\}$ and let $h_0$ correspond to the triple
$(i,j,k)$ under the bijection $\theta$ described in Section~\ref{subsection:reps}.
(Note that $h_0$ is necessarily a non-trivial element of $H$.)
We assume that we have gates ${\cal F}_X^{\phantom{\dagger}}$ and 
${\cal F}_H^{\phantom{\dagger}}$ 
at our disposal to perform the quantum Fourier transforms on $X=\ZZ_n$ and $H$.
We use the quantum circuit in Figure~\ref{fig:groupcircuit}.
This operates as follows.
\begin{eqnarray*}
\ket{\idgp_X,h_0} & %\stackrel{{\cal F}_G^{\phantom{\dagger}}\tens 
%{\cal F}_H^{\phantom{\dagger}}}{\mapsto} 
\xrightarrow{{\cal F}_X^{\phantom{\dagger}}\tens {\cal
    F}_H^{\phantom{\dagger}}} &
\sum_{g\in X,h\in H}\tau^{\idgp_X}_X(g)\tau_H^{h_0}(h)\ket{g,h} \\
& = & \frac{1}{\sqrt{|X|}}\sum_{g\in X,h\in H}\tau_H^{h_0}(h)\ket{g,h}, 
\mbox{ using~(\ref{eqn:tauidentity}),}\\
& \xrightarrow{U_f}&  \frac{1}{\sqrt{|X|}}\sum_{g\in X,h\in H}
\tau_H^{h_0}(h)\ket{g,f(g)h} \\
& \xrightarrow{{\cal F}^{\dagger}_X\tens{\cal F}^{\dagger}_H} &
\frac{1}{\sqrt{|X|}}
\sum_{g,g\pmd\in X, h,h\pmd\in H}\tau_H^{h_0}(h)\overline{\tau_X^{g\pmd}(g)
\tau_H^{h\pmd}(f(g)h)}\ket{g\pmd,h\pmd}.
\end{eqnarray*}
Given that $\theta$ is a bijection from $H$ to $R$, with
$\theta(h_0)=(i,j,k)$ we may sum over triples $(r,s,t)\in R$ 
instead of $h\pmd\in H$. The expression above then becomes
\[ \frac{1}{\sqrt{|X|}}\sum_{g,g\pmd\in X,h\in H,r,s,t}
\tau_{i,j}^k(h)\overline{\tau_X^{g\pmd}(g)\tau_{r,s}^t(f(g)h)}
\ket{g\pmd,h\pmd}.
\]
Applying Lemma~\ref{lemma:matrixmult}, this is equal to 
\begin{align} \label{equation:starstar}
\frac{1}{\sqrt{|X|}}\sum_{g,g\pmd\in X,r,s,t}
\overline{\tau_X^{g\pmd}(g)\tau_{r,i}^k(f(g))}\delta_{j,s}\delta_{k,t}\ket{g\pmd,(r,s,t)}\nonumber\\
=\frac{1}{\sqrt{|X|}}\sum_{g,g\pmd\in X,r=1,\ldots,n}
\overline{\tau_X^{g\pmd}(g)\tau_{r,i}^k(f(g))}\ket{g\pmd,(r,j,k)}.
\end{align}
Restricting to $g\pmd=\idgp_X$ on  the right hand side of
equation~(\ref{equation:starstar}) we obtain
\begin{equation} \label{equation:e_Gcase}
\frac{1}{|X|}\sum_{g \in X,r=1,\ldots,n}\overline{\tau_{r,i}^k(f(g))}\ket{\idgp_X,(r,j,k)}.
\end{equation}
If $f$ is $\rho_i^k$-balanced then  we have
$\sum_{g\in X}\overline{\tau_{r,i}^k(f(g))}=0$,  for 
$r=1,\ldots,n$; so (\ref{equation:e_Gcase}) is equal to
$0$. Thus measurement of  the first register never results in $\ket{\idgp_X}$. On the other hand,
if $f$ is $\rho_i^k$-constant then  there exists
a non-zero complex constant $c_r$ such that we have
$\tau_{i,r}^k(f(g))=c_r$, for all $g\in X$, for $r=1,\ldots,n$.
In this case  the right hand side of equation~(\ref{equation:starstar}) becomes
\[ \sum_{r=1}^m \overline{c_r}\ket{\idgp_X,(r,j,k)}
\]
and measurement of the first register always results in $\ket{\idgp_X}$.
\end{proof}

\section{Conclusion}\label{section:cases}

From Theorem \ref{theo:main} it follows that we can distinguish in a single step, 
with certainty, 
between $\rho^k_i$-constant
and $\rho^k_i$-balanced functions in all the examples of Section \ref{section:ab}.
In particular, for an Abelian group $A$ this means 
we may distinguish between $k$-constant and $k$-balanced
functions, for all $k\in A$, as described in Section \ref{sub:abelian}.
In the case of non-Abelian groups, as shown in Section \ref{sub:nonabelian},
there are many functions which may fall into the category of $\rho^k_i$-constant
or $\rho^k_i$-balanced for an appropriate choice of representation $\rho$.  
Here we summarise various known algorithms which are also covered by 
Theorem \ref{theo:main}.

\begin{enumerate}
\item\label{item:hoyer}
{\bf The Deutsch--Jozsa--H\o yer Algorithm:}
\bl
Suppose that $X$ and $H$ are finite groups with $H$ nontrivial 
such that $|X|=m|H|$ and assume that $f$ is constant or $m$-to-one (the
second possibility is called perfectly balanced in \cite{hoyer2}). If $f$ is 
constant then it is $\chi$-constant for any linear character $\chi$ of $H$. 
Suppose that $f$ is $m$-to-one and $\chi$ is a nontrivial linear character of $H$.
Let $\chi_0$ be the trivial character of $H$.
Then we have
\begin{eqnarray*}
\sum_{g\pmd\in X}\chi(f(g\pmd)) & = & m\sum_{h\pmd\in H}\chi(h\pmd) \\
& = & m\ip{\chi,\chi_0} \\
& = & 0,
\end{eqnarray*}
by orthogonality of irreducible characters, since $\chi$ is non-trivial.
So $f$ is $\chi$-balanced. Thus we recover Hoyer's result from \cite{hoyer2} that we can
distinguish between perfectly balanced and constant functions from $X$ to $H$
with certainty in a single quantum query.
\item {\bf The Deutsch--Jozsa--Constantini--Smeraldi Algorithm:}
\bl
In the case where $X=\ZZ_{mn}$ and $H=\ZZ_n$ we recover the result of \cite{constantinismeraldi},
which is itself a subcase of Hoyer's result \ref{item:hoyer}. 
\item {\bf The Deutsch Algorithm:}
\bl
In the case where $X=\ZZ_2$ and $H=\ZZ_2$ we obtain Deutsch's algorithm \cite{deutsch}.
\item {\bf Limited Surjectivity Testing:}
\bl
Suppose that $f:\ZZ_p\ra\ZZ_p$ where $p$ is prime. 
If we are promised that $f$ is either constant
or surjective then we can decide which is the case in a single quantum query, by 
Corollaries~\ref{corollary:primeconstant} and~\ref{corollary:primebalanced}. 
Classically, we would clearly require two queries.
(This is also a special case of the Constantini-Smeraldi result above.)
\item {\bf The Deutsch--Jozsa algorithm:}
\bl
In the case where $X=\ZZ_2^n$ and $H=\ZZ_2$ we obtain the Deutsch--Jozsa algorithm \cite{deutschjozsa},
in the form it appears in in \cite{cleveetal}.
%%%%%%%%%%%%%%%%%%%%%%%%%%
\begin{comment}
\item {\bf A More General Case Which Also Reduces to the Deutsch--Jozsa Algorithm:}
\bl
Let $G=H=\ZZ_2^n$. Since $H$ is Abelian, $\hat{H}$ is canonically isomorphic to
$H$. Thus, for $h_0\in H$, we can speak of ``$h_0$-balanced'' meaning $h_0$-balanced,
where $h_0$ corresponds to $h_0$ under this isomorphism. 
If $h_0=m$ is a power of $2$ then $f:G\ra G$ is $m$-constant/balanced if and only
if the function $G\ra\ZZ_2$ to the $m^{\rm th}$ qubit is $m$-constant/balanced. Suppose that
$f$ is either constant, $2^{n-1}$-to-one or that for some $m\pmd$ whose binary expression
contains exactly two $1$s. By letting $m$ vary over all powers of $2$ and $m\pmd$
we can decide in $m+1$ queries whether or not $f$ is constant, $2^{n-1}-to-one$ or
{\em balanced with separator }$m\pmd$ which means that $f$ is balanced for all the particular
values of $m$.
\notes{MB: This is the content of [BB].
As an anonymous referee has pointed out ...}
This can also be done more simply by $m+1$
applications of the standard Deutsch--Jozsa algorithm applied to $f$ restricted to each separate 
qubit, and then the sum of $f$ on two qubits.

\notes{MB: What we need to use this method effectively is to do it in a situation where there is 
no direct product structure and it doesn't decompose into qubits nicely. Such a situation might
be in general $p$-groups.}
\end{comment}
\end{enumerate}

These examples cover the main instances of the Deutsch--Jozsa--H\o yer algorithm,
of which we are aware, and in which the circuit is used to give an exact result.
Moreover the examples of Section \ref{section:ab} cover much wider classes of
functions than those covered by
the examples described in this section.
It therefore seems that Theorem \ref{theo:main} is a genuine generalisation of 
the algorithms existing in the literature. 

\section{Appendix 1}\label{section:appendix0}
In this section we provide a short summary of the standard properties of representations of finite
groups. Proofs may be found in any introductory text book on representation theory,
for example \cite{ledermann} or \cite{fultonharris}.

A {\em representation} of a group $G$ is a homomorphism $\rho:G\ra \GL(n,\co)$ for 
some $n\in\nat$. Given a basis $B$ of $\co^n$ and 
an element $g\in G$ we denote by $\rho_B(g)$ the matrix of the linear transformation
$\rho(g)$ with respect to the basis $B$. 
(If $B$ is understood we use $\rho(g)$ for both the linear transformation and its
matrix.)  
Suppose that $\rho\pmd$ is another representation of $G$ and there exists a matrix 
$T\in\GL(n,\co)$ such that $\rho=T^{-1}\rho\pmd T$. Then these representations are
not equal because they are different homomorphisms. However, there exists
a basis $B\pmd$ of $\co^n$ ($T$ is the change of basis matrix from $B$ to $B\pmd$)
such that for all $g\in G$, $\rho\pmd_{B\pmd}(g)=\rho_B(g)$. In this case we say that $\rho$ and
$\rho\pmd$ are {\em equivalent} representations.

If there is a proper subspace $V$ of $\co^n$ which is invariant under the action of $\rho(g)$ for all $g\in G$ (i.e. for all $g\in G$ we have $[\rho(g)](V)=V$) then $\rho$ is equivalent to a
direct sum $\rho_1\oplus\rho_2$ of smaller dimensional representations $\rho_1$ and $\rho_2$. 
If there is no such subspace then we say that $\rho$ is {\em irreducible}.

A  group $G$ always has the one-dimensional representation 
$\rho^1:G\ra\co$ given by $\rho^1(g)=1$ for all $g\in G$. This is called the
{\em trivial} representation of $G$ and is clearly irreducible.
Let $\co G$ be the vector space spanned by the elements of $G$. In the case where $G$ is finite, 
this is of course finite-dimensional. $G$ acts on itself by left (or right) multiplication
and this action extends to a linear map of $\co G$ to itself by permuting the vectors in its $G$-basis.
This is known as the left (or right) {\em regular representation} of $G$. The regular 
representation is not irreducible unless $G$ is trivial (see \cite{ledermann} Section 2.2). Furthermore,  the regular representation of a 
finite group $G$ decomposes as a direct sum of all of the (inequivalent) irreducible 
representations $\rho$ of $G$, each one appearing $\dim(\rho)$ times in the decomposition. 
It follows that (a) there are only finitely many irreducible representations of $G$ and (b) the
sum of the squares of the dimensions of the irreducible representations is equal to $|G|$.

If $\rho$ is a representation of $G$ such that for all $g\in G$, $\rho(g)$ is a unitary
map, then $\rho$ is called a {\em unitary representation} of $G$. If $G$ is a finite group then
a technique known as ``Weyl's unitary trick'' can be used to unitarize any irreducible representation (i.e. find an equivalent representation which is unitary.) That we can do this
 is important for the 
definition of the Fourier transform on $G$ so we recall its proof from \cite{terras}.
Let $\rho$ be an irreducible representation of $G$ with $n=\dim\rho$ and let 
$\langle \cdot,\cdot \rangle$ denote the standard inner product on
$\co^n$. First, we form an inner product $\langle \cdot,\cdot \rangle_{\rm inv}$ 
on $\co^n$ which is invariant under $\rho(g)$ for
all $g\in G$. This is done simply by defining
\[ \langle u,v \rangle_{\rm inv}=\sum_{g\in G}\langle \rho(g)u,\rho(g)v \rangle. 
\]
We show that, since the latter  inner product is invariant under 
$\rho(g)$ for all $g\in G$,  
$\rho$ is conjugate to a unitary representation. Suppose that $\{e_i\}_{i=1}^n$ is the
standard basis of $\co^n$. Let $C=\{c_{i,j}\}$ be the matrix
given by $c_{i,j}=\langle e_i,e_j \rangle_{\rm inv}$. Then $C$ is positive definite and
Hermitian ($C^\ast=C$). Thus, by the spectral theorem for Hermitian matrices, 
$C=U^\ast D U$ where $U$ is unitary and $D$ is diagonal with
positive real entries. Thus we can define $\sqrt D$ to be the matrix with entries the square
roots of the diagonal entries of $D$. Let $R=U^\ast \sqrt D U$. Then $C=R^2$ and since the inner
product which gave rise to $C$ is invariant under $\rho$ we have $\rho(g)C[\rho(g)]^\ast=C$ for
all $g\in G$. Let $\rho_U(g)=R^{-1}\rho(g)R$. We claim that $\rho_U$ is a unitary 
representation. This follows because for all $g\in G$ we have
\begin{eqnarray*}
\rho_U(g)[\rho_U(g)]^\ast & = & R^{-1}\rho(g)RR^\ast [\rho(g)]^{\ast}(R^{-1})^\ast \\
& = & R^{-1} \rho(g) R^2 [\rho(g)]^\ast (R^{-1})^{\ast}, \mbox { since $R$ is also Hermitian} \\
& = & R^{-1}\rho(g)C[\rho(g)]^\ast(R^{-1})^\ast \\
& = & R^{-1}C(R^{-1})^\ast \\
& = & R^{-1}RR^\ast (R^\ast)^{-1} \\
& = & I
\end{eqnarray*}
and similarly, for all $g$ in $G$, $[\rho_U(g)]^\ast\rho_U(g)=I$.

\section{Appendix 2}\label{section:appendix}
Here we recall the definition and some of the basic properties of
cyclotomic polynomials and establish the identity that we require in
Section \ref{section:ab}. Let $n$ be a positive integer, 
let $\omega = \exp(2\pi i/n)$ and let $R=\{d:d\in\ZZ, 1\le d<n,
\gcd(d,n)=1\}$ . The $n$th 
cyclotomic polynomial is defined to be 
\[\Phi_n(x)=\prod_{d\in R} (x-\omega^d).\]
It follows from the definition that the degree of $\Phi_n$ is
$\phi(n)$, where $\phi$ is Euler's totient function. 
As shown in, for example, \cite[p.194]{irelandrosen} 
\[x^n-1=\prod_{d|n}\Phi_d(x),\]
from which it follows that $\Phi_n\in \ZZ[x]$. Moreover (see \cite{irelandrosen}
$\Phi_n$ is irreducible in $\ZZ[x]$ and so is 
the minimum polynomial of $w$ over $\QQ$.
%%%%%%%%%%%%%%%%%%%%%%%%
The following identity is standard; we cast it in
the particular form we require below.
Let $n=p^\gamma s$, where $p$ is prime, $\gamma\ge 1$ and $p\nmid s$.
We have
\begin{align*}
x^n-1&=\prod_{d|n}\Phi_d(x)\\
&= \prod_{d|s}\Phi_{p^\gamma d}(x)\prod_{d|n/p}\Phi_d(x)\\
&=\prod_{d|s}\Phi_{p^\gamma d}(x) (x^{n/p}-1).
\end{align*}
As $x^n-1=(x^{n/p}-1)\Phi_p(x^{n/p})$ it follows that
\begin{equation}\label{eq:Fr}
\Phi_p(x^{n/p})=\prod_{d|s}\Phi_{p^\gamma d}(x).
\end{equation}
%%%%%%%%%%%%%%%%%%%%%%%%%%%%%

We shall also require the following fact.
\begin{lemma}
Let $a_1,\ldots,a_n$ be non-negative integers, $n\ge 2$, and let
$d=\gcd(a_1+1,\ldots ,a_n+1)$. Then there exist polynomials $s_i(x)\in
\ZZ[x]$, $i=1,\ldots,n$,  such that
$$\sum_{i=1}^n s_i(x)(1+x+\cdots +x^{a_i})=1+x+\cdots+x^{d-1}.$$ 
\end{lemma}
\begin{proof}
First consider the case $n=2$. Note that if $a_1=a_2$ then $d=a_1+1$ and
$s_1=1$, $s_2=0$ have the required property. 
Use induction on $a_1+a_2$, 
starting with the case $a_1=a_2=0$. In this case 
the result follows from the previous remark.
Now suppose that
$a_1+a_2>0$. It may be assumed that $a_1<a_2$. Set
$f=1+x+\cdots +x^{a_1}$, $g=1+x+\cdots +x^{a_2}$,
$t=-x^{a_2-a_1}$ and $h=1+x+\cdots +x^{a_2-a_1-1}=tf+g$.

Then
$\gcd(a_1+1,(a_2-a_1-1)+1)=\gcd(a_1+1,a_2+1-(a_1+1))=\gcd(a_1+1,a_2+1)=d,$
so by induction there exist $s^\prime_1,s^\prime_2$ such that 
$s^\prime_1 f+s^\prime_2 h=1+x+\cdots +x^{d-1}.$
Hence 
\begin{align*}
1+x+\cdots +x^{d-1} &= s^\prime_1 f+s^\prime_2 (tf+g)\\
&=  (s^\prime_1 +s^\prime_2 t)f+s^\prime_2 g,
\end{align*}
as required. Thus the result holds when $n=2$.

Now suppose that $n>2$. Let $d_1=\gcd(a_1+1,\ldots ,a_{n-1}+1)$ and 
$d_2=\gcd(a_{n-1}+1,a_n+1)$. From the inductive hypothesis there
exist $u_1,\ldots ,u_{n-1},v_1,v_2\in \ZZ[x]$ such that 
\begin{equation}
\sum_{1}^{n-1} u_i(x)(1+x+\cdots +x^{a_i})=1+x+\cdots
+x^{d_1-1}\label{eq:gcd1}
\end{equation}
and
\begin{equation}
v_1(x)(1+x+\cdots +x^{a_{n-1}})+v_2(x)(1+x+\cdots +x^{a_n})=1+x+\cdots
+x^{d_2-1}.\label{eq:gcd2}
\end{equation}
As $d=\gcd(d_1,d_2)$ there are $u,v\in \ZZ[x]$ such that 
\begin{equation}
u(x)(1+x+\cdots +x^{d_1-1})+v(x)(1+x+\cdots +x^{d_2-1})=1+x+\cdots
+x^{d-1}.\label{eq:gcd3}
\end{equation}
Combining \eqref{eq:gcd1}, \eqref{eq:gcd2} and \eqref{eq:gcd3} gives
the required result.
\end{proof}

Define $F_n(x)=1+x+\cdots +x^{n-1}$, for all integers $n\ge 1$.
\begin{corollary}\label{cor:pol}
If $p_1,\ldots ,p_n$ are distinct primes set $m=p_1\cdots p_n$ and
$m_i=m/p_i$. Then there exist $s_1,\ldots ,s_n\in\ZZ[x]$ such that 
$$\sum_1^n s_iF_{m_i}=1.$$
\end{corollary}
%\begin{proof}
%This follows from the lemma, since $\gcd(m_1,\ldots ,m_n)=1$.
%\end{proof}
\begin{theo}\label{theo:dephi}
Let $n=p^\alpha q^\beta$ be a positive integer, 
where $p$ and $q$ are distinct primes and $\alpha$ and $\beta$ are
non-negative integers.
Let $g\in\ZZ[x]$ such that $\Phi_n(x)|g(x)$, deg$(g)=n-1$ and the
coefficients of $g$ are all non-negative. 
\begin{enumerate}
\item If $\alpha\ge 1$ and $\beta =0$ then
\begin{equation}\label{eq:gphia}
  g(x)=s(x)\Phi_p(x^{n/p}),
\end{equation} 
for some $s\in\ZZ[x]$ with
  non-negative coefficients.
\item If $\alpha\ge 1$ and $\beta\ge 1$ then there exist $s_1,
s_2\in\ZZ[x]$ such that 
\begin{equation}\label{eq:gphi}
g(x)=s_1(x)\Phi_p(x^{n/p})+s_2(x)\Phi_q(x^{n/q}),
\end{equation}
and the coefficients of $s_1$ and $s_2$ are all non-negative. 
\end{enumerate}
\end{theo}
\begin{proof}
We begin by proving that there exist elements $s$ or $s_i$ in $\ZZ[x]$ such that
\eqref{eq:gphia} or \eqref{eq:gphi}
holds, as appropriate, and subsequently show that $s$ or the $s_i$ may be chosen so that their 
coefficients are non-negative.

If $n=p^\alpha$ then \eqref{eq:Fr} yields $\Phi_n(x)=\Phi_p(x^{n/p})$,
so we may write $g(x)=s_p(x)\Phi_n(x)=s_p(x)\Phi_p(x^{n/p})$, with
$s_p\in \ZZ[x]$, as required. Assume then that $\alpha\ge 1$ and $\beta\ge 1$.
As $\Phi_r=F_r$ when $r$ is prime, we have
\begin{equation}\label{eq:Fp}
F_p(x^{n/p})=\prod_{d|q^\beta}\Phi_{p^\alpha d}(x)
\end{equation}
and
\begin{equation}\label{eq:Fq}
F_q(x^{n/q})=\prod_{d|p^\alpha}\Phi_{q^\beta d}(x).
\end{equation}

Write 
\begin{equation}\label{eq:gf}
g(x)=f(x)\Phi_n(x), \textrm{ where } f\in\ZZ[x].
\end{equation}
From  Corollary \ref{cor:pol} there are polynomials $s_1$ and $s_2\in
\ZZ[x]$ such that 
$1=s_1(x)F_p(x)+s_2(x)F_q(x)$. Let $k=n/pq$ and replace $x$ with $x^k$
in the previous equality to obtain
$1=s_1(x^k)F_p(x^k)+s_2(x^k)F_q(x^k)$.
Multiplying through by $f(x)$ gives 
$f(x)=s_p(x)F_p(x^k)+s_q(x)F_q(x^k),$ for some $s_p,s_q\in\ZZ[x].$
Hence
$g(x)=(s_p(x)F_p(x^k)+s_q(x)F_q(x^k))\Phi_n(x).$ Applying
\eqref{eq:Fp} and \eqref{eq:Fq}, with $n/q$ and $n/p$ in place of
$n$, respectively, we
obtain
\begin{align*}
g(x)&=\left(s_p(x)\prod_{d|q^{\beta -1}}\Phi_{p^\alpha
    d}(x)+s_q(x)\prod_{d|p^{\alpha -1}}\Phi_{q^\beta
    d}(x)\right)\Phi_n\\
&=s_p(x)\prod_{d|q^\beta}\Phi_{p^\alpha d}(x) +
s_q(x)\prod_{d|p^\alpha d}\Phi_{q^\beta d}(x)\\
&=s_p(x)F_p(x^{n/p})+s_q(x)F_q(x^{n/q}),
\end{align*}
using \eqref{eq:Fp} and \eqref{eq:Fq} again. Thus we have $s_p$ and
$s_q\in \ZZ[x]$ as required. Next we shall show that $s_p$ and $s_q$
may be chosen so that their coefficients are all non-negative.

Note that as deg$(g)=n-1$ and
deg$(F_r)=r-1$ we have deg$(s_p)\le n/p-1$ and deg$(s_q)\le n/q-1$. Let
$n_p=n/p$ and $n_q=n/q$. Then we may write 
$$s_p(x)=u_0+u_1 x+ \cdots +u_{n_p-1}x^{n_p-1} \textrm{ and }
s_q(x)=v_0 +v_1 x +\cdots +v_{n_q-1}x^{n_q-1},$$
for suitable $u_i$ and $v_i\in\ZZ$. Set $A_p(x)=s_p(x)F_p(x^{n_p})$
and $A_q(x)=s_q(x)F_q(x^{n_q})$. If $0\le r<n$ then the coefficient of
$x^r$ in $A_p(x)$ is $\sum u_j$, where the sum runs over those $j$
such that $0\le j<n_p$ and $j+k n_p=r$, for some $k\in \ZZ$. There is
a unique pair $(k,j)$ with this property, for each such $r$. Hence the
coefficient of $x^r$ in $A_p(x)$ is $u_j$, where $j$ is the unique
integer such that $0\le j<n_p$ and $r\equiv j (\mod n_p)$. 
If $\beta=0$ then $g(x)=A_p(x)$ and  it follows that the $u_j$'s are all non-negative and
the result follows with $s=s_p$. 
Assume from now on that $\beta\ge 1$. Then 
the coefficient of  $x^r$ in $A_q(x)$ is $u_l$, where $l$ is the
unique integer such that $0\le l<n_q$ and $r\equiv l(\mod n_q)$.

Let $d=\gcd(n_p,n_q)=p^{\alpha -1}q^{\beta-1}$. For $j=0,\ldots ,d-1$ define 
$$s_{p,j}(x)=\sum_{i=0}^{q-1}u_{id+j}x^{id+j}$$
and 
$$s_{q,j}(x)=\sum_{i=0}^{p-1}v_{id+j}x^{id+j}.$$
Then $$s_p(x)=\sum_{j=0}^{d-1}s_{p,j}(x)\textrm{ and }
s_q(x)=\sum_{j=0}^{d-1}s_{q,j}(x).$$
Fix $j$ and $a$ with $0\le j<d$ and $0\le a<q$. If $0\le l<n_q$ then
there exists $r$ such that $0\le r<n$ with $r\equiv ad+j (\mod n_p)$
and $r\equiv l (\mod n_q)$ if and only if $ad+j\equiv l(\mod d)$
(using the Chinese Remainder Theorem) if and only if $l=bd+j$, for
some $b\in\ZZ$. Moreover, $0\le b<p$, as $0\le l<n_q$.

Therefore, for all $j,a,b$ with $0\le j<d$, $0\le a<q$ and $0\le b<p$,
there exists  an integer $r=r(j,a,b)$, unique modulo $n$, such that
\begin{equation}\label{eq:nr}
n_r=u_{ad+j}+v_{bd+j}.
\end{equation}
Conversely, if $0\le r<n$ then there exist $j,a,b$ in the above
ranges, such that \eqref{eq:nr} holds. For fixed $j$ with $0\le j<d$
define 
$$c_j=\min\{u_{ad+j},v_{bd+j}:0\le a<q,0\le b<p\}.$$
If $c_j\ge 0$ define $t_{p,j}=s_{p,j}$ and $t_{q,j}=s_{q,j}$. If
$c_j<0$ then there is some $a$ or $b$ such that $c_j=u_{ad+j}$ or
$c_j=v_{bd+j}$. Suppose that $c_j=u_{ad+j}<0$. Since $n_r\ge0$, for 
$r=0,\ldots,n-1$, it follows from \eqref{eq:nr} that 
$u_{ad+j}+v_{bd+j}\ge 0$ and so $v_{bd+j}\ge |c_j|$, for $b=0,\ldots
,p-1$. By definition, $u_{id+j}\ge c_j,$ for $i=0,\ldots ,q-1$, so
setting
$$t_{p,j}(x)=s_{p,j}(x)+|c_j|x^j F_q(x^d)$$
and 
$$t_{q,j}(x)=s_{q,j}(x)-|c_j|x^j F_p(x^d)$$
the polynomials $t_{p,j}$ and $t_{q,j}$ have non-negative integer
coefficients. If $c_j\neq u_{ad+j}$ but $c_j=v_{bd+j}$, for some $b$,
we construct $t_{p,j}$ and $t_{q,j}$ in the same way, reversing the
roles of $p$ and $q$, and obtain the same result.

Now fix $j$ such that $c_j<0$. Assume that $c_j=u_{ad+j}$. Then
\begin{align}
t_{p,j}(x)F_p(x^{n/p})+t_{q,j}(x)F_q(x^{n/q})
&=s_{p,j}(x)F_p(x^{n/p})+s_{q,j}(x)F_q(x^{n/q})\nonumber\\
&+|c_j|x^j(F_q(x^d)F_p(x^{n/p})-F_p(x^d)F_q(x^{n/q}))\label{eq:teq}
\end{align}
We have
\begin{align*}
F_q(x^d)F_p(x^{n/p})&=F_q(x^{p^{\alpha -1}q^{\beta -1}})F_p(x^{p^{\alpha
      -1}q^\beta})\\
&= \prod_{d|p^{\alpha-1}}\Phi_{q^\beta
    d}(x)\prod_{e|q^\beta}\Phi_{p^\alpha e}(x), \textrm{ from
    \eqref{eq:Fp} and \eqref{eq:Fq}},\\
&=\left( \prod_{d|p^{\alpha-1}}\Phi_{q^\beta
    d}(x)\prod_{e|q^{\beta -1}}\Phi_{p^\alpha
    e}(x)\right)\Phi_{p^\alpha q^\beta}(x)\\
&=\prod_{d|p^{\alpha}}\Phi_{q^\beta
    d}(x)\prod_{e|q^{\beta -1}}\Phi_{p^\alpha
    e}(x)\\
&=F_q(x^{n/q})F_p(x^d).
\end{align*}
Therefore, from \eqref{eq:teq}
\begin{align}
t_{p,j}(x)F_p(x^{n/p})+t_{q,j}(x)F_q(x^{n/q}) &=
s_{p,j}(x)F_p(x^{n/p})+s_{q,j}(x)F_q(x^{n/q}).
\label{eq:ts}
\end{align}
Now define $$s_1(x)=\sum_{j=0}^{d-1} t_{p,j}(x)$$
and $$s_2(x)=\sum_{j=0}^{d-1} t_{q,j}(x).$$
Then the coefficients of $s_1$ and $s_2$ are non-negative and 
it follows from \eqref{eq:ts} that 
$$g(x)=s_p(x)F_p(x^{n/p})+s_q(x)F_q(x^{n/q})=s_1(x)F_p(x^{n/p})+s_2(x)F_q(x^{n/q}),$$
as required.
\end{proof}
\subsubsection*{Acknowledgements}
The authors are grateful to H. Buhrman, S. Linton, M. Mosca and C. Smyth for
useful contributions and helpful suggestions in writing this article.

{\small This research was supported by EPSRC MathFIT grant number GR/87406.}
\bl
{\small Samuel L. Braunstein currently holds a Royal Society Wolfson Research Merit Award.}
\end{document}